\documentclass[reprint,amsmath,amssymb,superscriptaddress,nolongbibliography,aps,prl]{revtex4-2}

\usepackage{xr}
\usepackage[utf8]{inputenc}
\usepackage{amsmath}
\usepackage{amsmath,amssymb}
\usepackage{siunitx}
\usepackage[dvipsnames]{xcolor}
\usepackage{color,soul}
\setstcolor{red}
\sethlcolor{yellow}
\usepackage{notes2bib}
\usepackage[colorlinks=true]{hyperref}
\hypersetup{
    allcolors = {blue},
}

\newcommand{\bd}[1]{{\boldsymbol{#1}}}
\newenvironment{eqs}%
{\begin{equation} \begin{aligned}}%
{\end{aligned} \end{equation} }
\newcommand{\beal}{\begin{eqs}}
\newcommand{\eal}{\end{eqs}}

\usepackage{graphicx}
\usepackage{dcolumn}% Align table columns on decimal point
\usepackage{bm}% bold math
\usepackage{booktabs}
\usepackage{framed,enumitem}

\begin{document}
\preprint{APS/123-QED}

\title{Bending Stiffness Collapse, Buckling, Topological Bands of Freestanding
Twisted Bilayer Graphene}
\author{Jin Wang$^{\dagger}$}
    \affiliation{International School for Advanced Studies (SISSA), Via Bonomea 265, 34136 Trieste, Italy}
    
\author{Ali Khosravi$^{\dagger}$}%
    \affiliation{International School for Advanced Studies (SISSA), Via Bonomea 265, 34136 Trieste, Italy}
    \affiliation{International Centre for Theoretical Physics (ICTP), Strada Costiera 11,34151 Trieste,Italy}
    \affiliation{CNR-IOM, Consiglio Nazionale delle Ricerche - Istituto Officina dei Materiali, c/o SISSA, Via Bonomea 265, 34136 Trieste, Italy}
    
\author{Andrea Silva}
    \affiliation{CNR-IOM, Consiglio Nazionale delle Ricerche - Istituto Officina dei Materiali, c/o SISSA, Via Bonomea 265, 34136 Trieste, Italy}%
    \affiliation{International School for Advanced Studies (SISSA), Via Bonomea 265, 34136 Trieste, Italy}%

\author{Michele Fabrizio}
    \affiliation{International School for Advanced Studies (SISSA), Via Bonomea 265, 34136 Trieste, Italy}

\author{Andrea Vanossi}
    \affiliation{CNR-IOM, Consiglio Nazionale delle Ricerche - Istituto Officina dei Materiali, c/o SISSA, Via Bonomea 265, 34136 Trieste, Italy}%
    \affiliation{International School for Advanced Studies (SISSA), Via Bonomea 265, 34136 Trieste, Italy}%

\author{Erio Tosatti}
  \email{tosatti@sissa.it}
  \affiliation{International School for Advanced Studies (SISSA), Via Bonomea 265, 34136 Trieste, Italy}
  \affiliation{International Centre for Theoretical Physics (ICTP), Strada Costiera 11,34151 Trieste,Italy}
  \affiliation{CNR-IOM, Consiglio Nazionale delle Ricerche - Istituto Officina dei Materiali, c/o SISSA, Via Bonomea 265, 34136 Trieste, Italy}

\date{\today}

\begin{abstract}
The freestanding  twisted  bilayer graphene (TBG) is unstable, below a critical twist angle $\theta_c \sim 3.7^\circ$, against a moir\'e $(2 \times 1) $ buckling distortion at $T=0$. Realistic simulations reveal the concurrent unexpected collapse of the bending rigidity, an unrelated macroscopic mechanical parameter. An analytical model connects bending and buckling anomalies at $T=0$, but as temperature rises the former fades, while buckling persists further.
The  $(2 \times 1) $ electronic properties are also surprising. The magic twist angle narrow bands, now eight in number, fail to show zone boundary splittings despite the new periodicity. Symmetry shows how this is dictated by an effective single valley physics. These structural, critical, and electronic predictions promise to make the freestanding state of TBG especially interesting.
\end{abstract}

\maketitle

\begin{figure*}[ht!]
\centering
\includegraphics[width=0.96\linewidth]{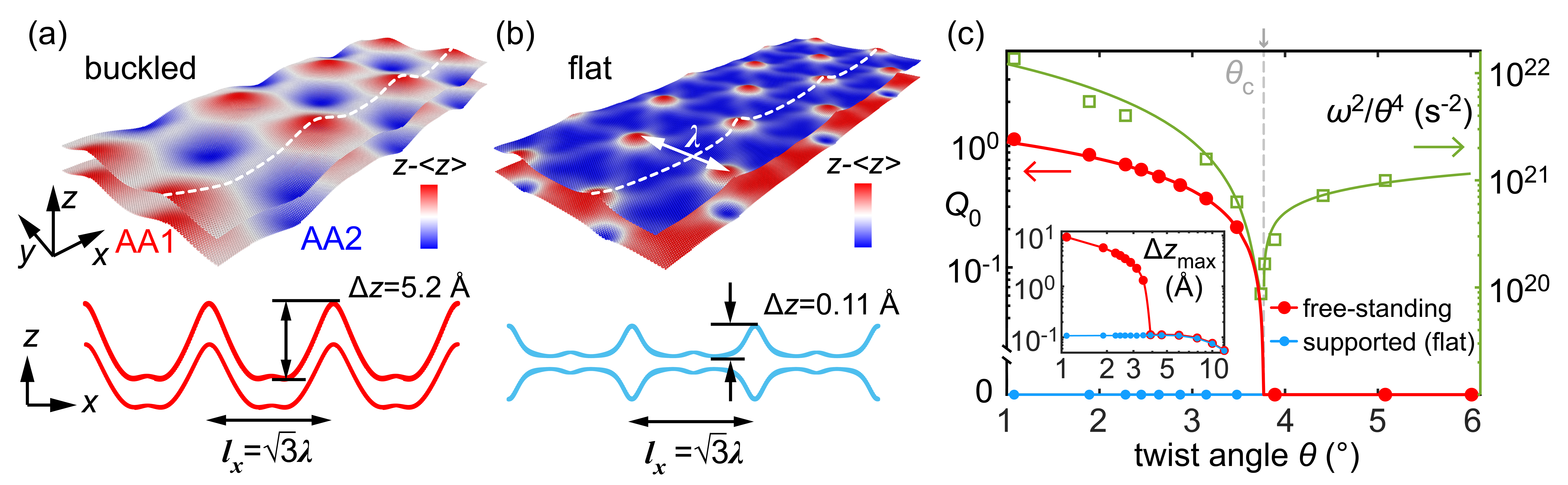}
\caption{Visual models of small twist angle TBG structure at $T=0$.
Out-of-plane displacement of (a) optimal free-standing $(2 \times 1)$ buckled and (b) flat structures, magnified by factors 3 and 100 respectively, are sketched for twist angle $\theta_m=1.08^\circ$, where the moir\'e size $\lambda \approx a_\mathrm{Gr}/\theta_m \approx 13~\textrm{nm}$. (c) Twist angle dependence of the 
optimal buckling order parameter $Q_0$ (red and blue, left axis) for $N_x =N_y =1$, $T=0$~K, and of the normalized
soft phonon frequencies $\omega^2/\theta^4$ (right axis, green).
Critical twist angle $\theta_c \approx 3.77^{\circ}$ marked by grey dashed line. Red and green lines are power law fits as described in text. Inset shows the maximum out-of-plane corrugation $\Delta z_{\rm max}$.
}
\label{fig:1}
\end{figure*}

Two-dimensional (2D) bilayers and multilayers with variable lattice mismatch and/or twist angles exhibit a host of physical properties that also hold promise for applications
\cite{Geim.nature.2013,Neto.science.2016,Yankowitz.NatRevPhys.2019,Liu.NatRevMater.2016}.
With exceptional electronic properties at the magic twist angle, twisted bilayers graphene (TBG) are prominent among them \cite{Bistritzer.PNAS.2011,Cao.nature.2018-1,Cao.nature.2018-2}.
Experimental bilayers are generally studied in deposited/encapsulated configurations, which preserve a flat geometry and the twist-related moir\'e pattern plays no mechanical role. Yet, TBG may also be realized as freestanding \cite{Butz.nature.2014, Ying.nc.2022}.
Below a critical twist angle $\theta_c$, moir\'e related structural instabilities and a variety of ``buckled" states were suggested by pioneering freestanding simulations \cite{Dai.nanolett.2016, Lamparski.2dmater.2020, Rakib.commphys.2022}, but
the actual nature and properties of the true TBG buckled state %and its novel properties 
remain unknown.
We use here theory and simulation to show
that the moir\'e buckled state formed at low twist angle is quite different from expectations. Mechanically, it  is accompanied by the unanticipated collapse of the TBG macroscopic bending rigidity. Electronically, the  magic twist angle narrow bands, now doubled in number, are unexpectedly degenerate at zone boundaries, the vanishing Bragg scattering symmetry motivated reflecting single valley physics.

Starting with molecular dynamics (MD) simulations, large size model TBGs with variable twist $\theta$ and variable numbers $N_\mathrm{moire}$ of moir\'e cells were constructed with periodic boundary conditions in the $(x,y)$ plane. Based on well-tested interatomic interactions and protocols (detailed in SI.I) we sought the zero stress equilibrium $T$=0 structure versus $\theta$. We found that, similar but not identical to suggestions \cite{Dai.nanolett.2016,Lamparski.2dmater.2020, Rakib.commphys.2022}, two regimes emerge, separated by a structural phase transition at a critical $\theta_c \approx 3.77 ^\circ$.
Above $\theta_c$ the two layers remain flat and specular relative to the central plane (Fig.~\ref{fig:1}b). Below $\theta_c$ the layers jointly buckle giving rise to a ``moir\'e ($2\times1$)" cell doubling along
armchair direction $x$, leading to two inequivalent, $z$-antisymmetrical 
AA nodes, AA1 (up) and AA2 (down) per cell, as in Fig.~\ref{fig:1}a. The magnitude of buckling is large. At the magic twist angle $\theta_m \approx 1.08^{\circ}$ for example the zigzag $z$-corrugation is $\approx  10 \mathrm{\AA}$ (Fig.~\ref{fig:1}a,c).
A competing ($\sqrt{3} \times \sqrt{3}$) buckling distortion, with one AA1 (100 \% up) hexagonally surrounded by AA2 and AA3 (50 \% down), was also found. It led to a slightly lower energy gain, and its details are not further pursued here (see however SI.VIII).

The energy gain driving the buckling distortion at $\theta < \theta_c$ is interlayer, with increased AB and BA, Bernal stacked areas, relative to the flat, unbuckled state's. That gain is balanced by an intralayer cost concentrated at the AA nodes, now transformed into buckling ``hinges" AA1 (up) and AA2 (down).
As $\theta$ decreases, the 2D density of AA nodes, thus of hinges, drops $\sim \theta^2$,
and so does the cost, eventually favoring buckling for $\theta \leq \theta_c$. The transition, simulated by maintaining zero external stress and zero temperature, was found to be continuous.
We define the $T=0$~K buckling order parameter $Q_0$ as the large-size average Fourier component of the $(2 \times 1)$ moir\'e corrugation
\begin{equation}
\begin{aligned}
        Q_0 &=\frac{a_\mathrm{Gr}}{N_x N_y A} \left \langle \sum\limits_{n=1}^{N_\mathrm{at}} z_n \exp\left(-\frac{2 \pi i}{l_x} x_n\right) \right \rangle
\end{aligned}
\label{eq:Q}
\end{equation}
where $a_\mathrm{Gr}$ is graphene's lattice constant,
$l_x=\sqrt{3} \lambda$ is the size of the buckled unit cell along the armchair direction $x$ (Fig.~\ref{fig:1}a,b), $\lambda \sim a_\mathrm{Gr}/\theta$ is the moir\'e lattice constant (Fig.~\ref{fig:1}a,b), $x_n$ and $z_n$ are coordinates of the $n$-th atom ($n=1, 2, ... N_\mathrm{at} $), $A= \sqrt{3} \lambda^2$ is the buckled unit cell area, and $N_x$, $N_y$ are the number of cell replicas along $x$ and $y$.
This $Q_0(\theta)$ is proportional to the buckling induced
bilayer thickness increase $\langle (z_{AA1} - z_{AA2}) \rangle $ (inset of Fig.~\ref{fig:1}c).
Simulation for a set of $\theta$ values (see SI.I) characterized by a sufficient numbers of moir\'e cells $N_\mathrm{moire}$
showed, at $T=0$, a growth of $Q_0(\theta < \theta_c$) (Fig.~\ref{fig:1}c) well approximated  by a power-law rise
$Q_0(\theta)  \approx 0.48  \Theta(\theta_c - \theta) (\theta_c - \theta)^{\beta}$, where $\Theta$ is Heaviside's function, and $\beta =0.7(0)$ a critical exponent. Within fitting uncertainty, reflected by the second decimal in brackets, this exponent differs from 1/2, that could be expected for a classical $T=0$ transition.

As in other displacive phase transitions, the local free energy around equilibrium supports a soft buckling phonon mode $\omega_{i}$ ($i=(+, -)$ refers to above or below $\theta_c$), a mode which will also control critical fluctuations at $T > 0$.
Its frequency was extracted from oscillations around equilibrium of a $(2 \times 1)$ moir\'e cell by starting the dynamics with $Q=Q_0 + \delta Q$, with $|\delta Q|/Q_0 \ll 1$, while maintaining $T=0$ and zero stress.
Because the cell area $A(\theta)$ varies as $\theta^{-2}$, it is convenient to further normalize the soft mode frequencies to constant area in the form $\omega^2/\theta^4$. Power law fits near the singularity at $\theta_c$ yield (Fig.~\ref{fig:1}c), $\omega^2_{i}/\theta^4 \sim a_i |\theta - \theta_c|^{\gamma_{i}}$, with $\gamma_{+} = 0.3(7)$, $\gamma_{-} = 1.3(0)$, with
$a_{+} = 8.58\times10^{20}~\mathrm{s}^{-2}$,
$a_{-} = 3.24\times10^{21}~\mathrm{s}^{-2}$.
Note the strongly asymmetric, again unusual exponents.
We observe nevertheless that $\gamma_{-}/\beta \approx 2$ as in standard mean-field theory.

\begin{figure}[hb!]
\centering
\includegraphics[width=1\linewidth]{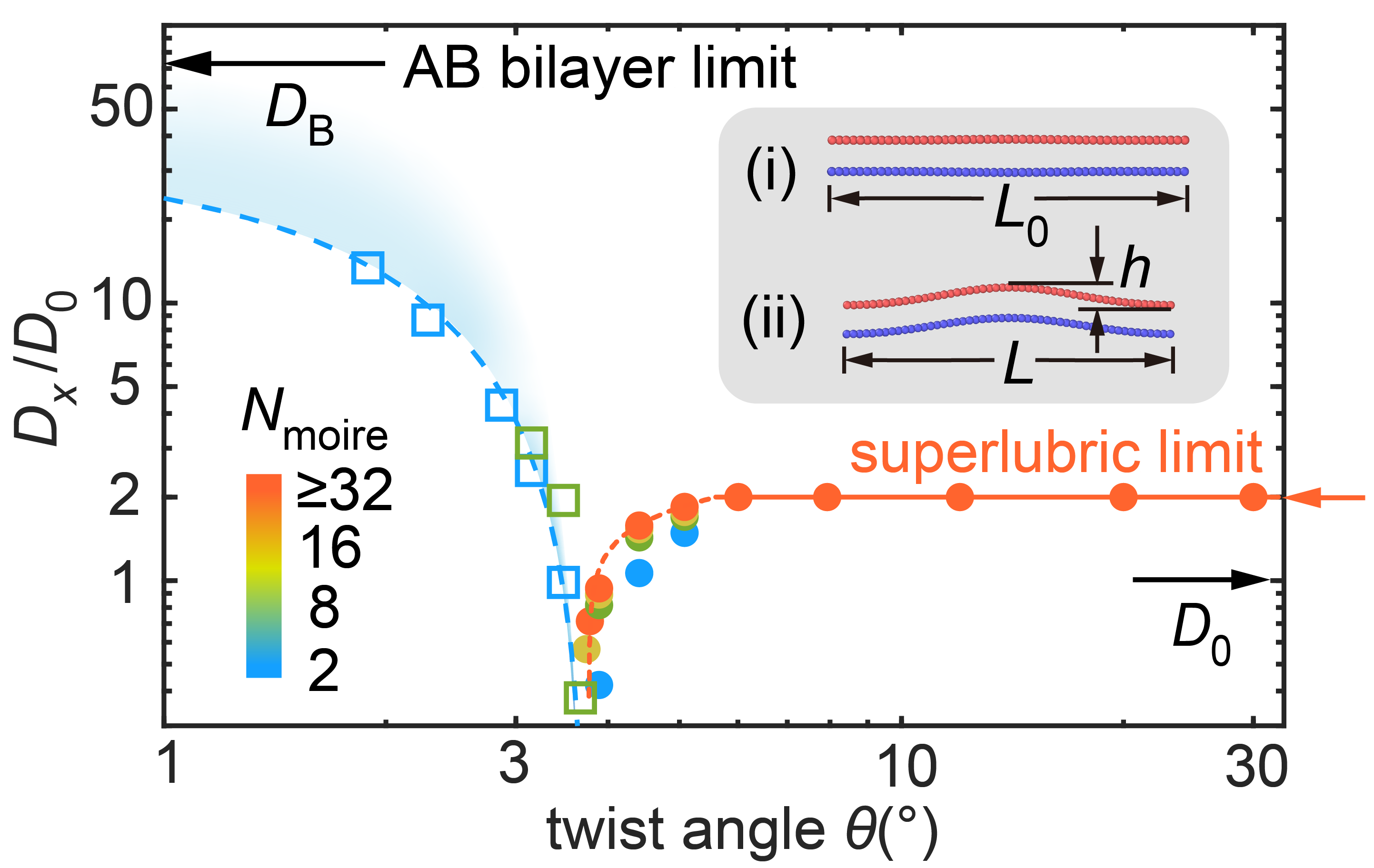}
\caption{Freestanding TBG bending stiffness $D_x/D_0 $, normalized to the monolayer's $D_0$, from zero stress simulations at  $T=0$.
Note the critical collapse (red circles for  $\theta > \theta_c$, blue squares for $\theta < \theta_c$). 
Dashed and dotted curves are power-law fits near $\theta_c$. Multiplicity of circles at same twist angle shows convergence for
simulation cells with increasing size $N_\mathrm{moire}$ (the $N_\mathrm{moire}$ dependence is weak for $\theta < \theta_c$). 
Inset: simulation protocol (see text).
}
\label{fig:2}
\end{figure}

More interestingly,
the buckling amplitude $Q_0$ and its soft mode frequency are not the only critical quantities at $\theta \to \theta_c$. We found an unexpected macroscopic partner, the bilayer's  bending stiffness.
Defined for direction $\mu = (x,y)$ as
$D_{\mu} = {\rm d} F/{\rm d}[(\partial^2 h/\partial \mu^2)^2]$,
where $F$ is the Helmholtz free energy density,
$\partial^2 h/\partial \mu^2$
is the $\mu$-th component of the 2D Laplacian, and $h$ the bilayer's corrugation profile $h(x,y)$ (Fig.~\ref{fig:2}). Controlling the membrane's deviations from planarity, $D_{\mu}$ 
determines the macroscopic flexural mode dispersion along $\mu$, $\omega_{\mu}(q_{\mu})=(D_{\mu}/\rho_\mathrm{2D})^{1/2} q_{\mu}^2$, of an infinite membrane of 2D density $\rho_\mathrm{2D}$.

The freestanding TBG bending stiffness $D_{\mu}$ was extracted by starting simulations with a slight $x$-compression (Fig.~\ref{fig:2}), i.e., $L_{\mu} = L_{0\mu} - \delta L$, of the bilayer's zero-pressure equilibrium size. Either the initial energy increase $\delta E$, or the ensuing  flexural oscillation $\omega_{\mu}$ yield $D_{\mu}= \lim_{\delta E \to 0} \frac{L_{\mu}^4 \delta E}{\pi^4 A h^2} =\frac{\rho_{\rm 2D}L_{\mu}^4 \omega_{\mu}^2}{16\pi^4}$ \cite{Landau.elasticity}.
The resulting $D_x$ is shown in Fig.~\ref{fig:2}. 
For $6^\circ<\theta< 30^\circ$, $D_x$ is close to $2 D_0$, where $D_0=1.44~\mathrm{eV}$ is the bending stiffness of monolayer graphene \cite{Lu.JphysD.2009}, the factor 2 reflecting free sliding between the two layers\cite{Han.natmat.2020, Yu.AdvMat.2021}.
At the opposite end $\theta=0^\circ$, layers are instead locked in Bernal's AB stacking. That pushes $D_x$ up to $D_\mathrm{B} \approx 100~\mathrm{eV}$, now reflecting the large in-plane stiffness of graphene \cite{Lee.nanolett.2012,Wang.prl.2019}.
The novelty is that between these extremes, $D_x$ drops below $2D_0$ when $ \theta \lessapprox  6^\circ$, critically collapsing at $\theta_c$, and rising immediately below towards $D_\mathrm{B}$. Near $\theta_c$ the collapse is critical
\begin{equation}
    D_x(\theta) \sim \mathrm{c}_{i} |\theta - \theta_c|^{\epsilon_i} 
\end{equation}
with exponents $\epsilon_{+}= 0.2(2) $ and $\epsilon_{-}= 1.4(4) $, and
$\mathrm{c}_{+} $=2.4 eV and $\mathrm{c}_{-} $ =7.5 eV,
for $\theta > \theta_c$ and $\theta < \theta_c$ respectively.

\begin{figure}[ht!]
\centering
\includegraphics[width=\linewidth]{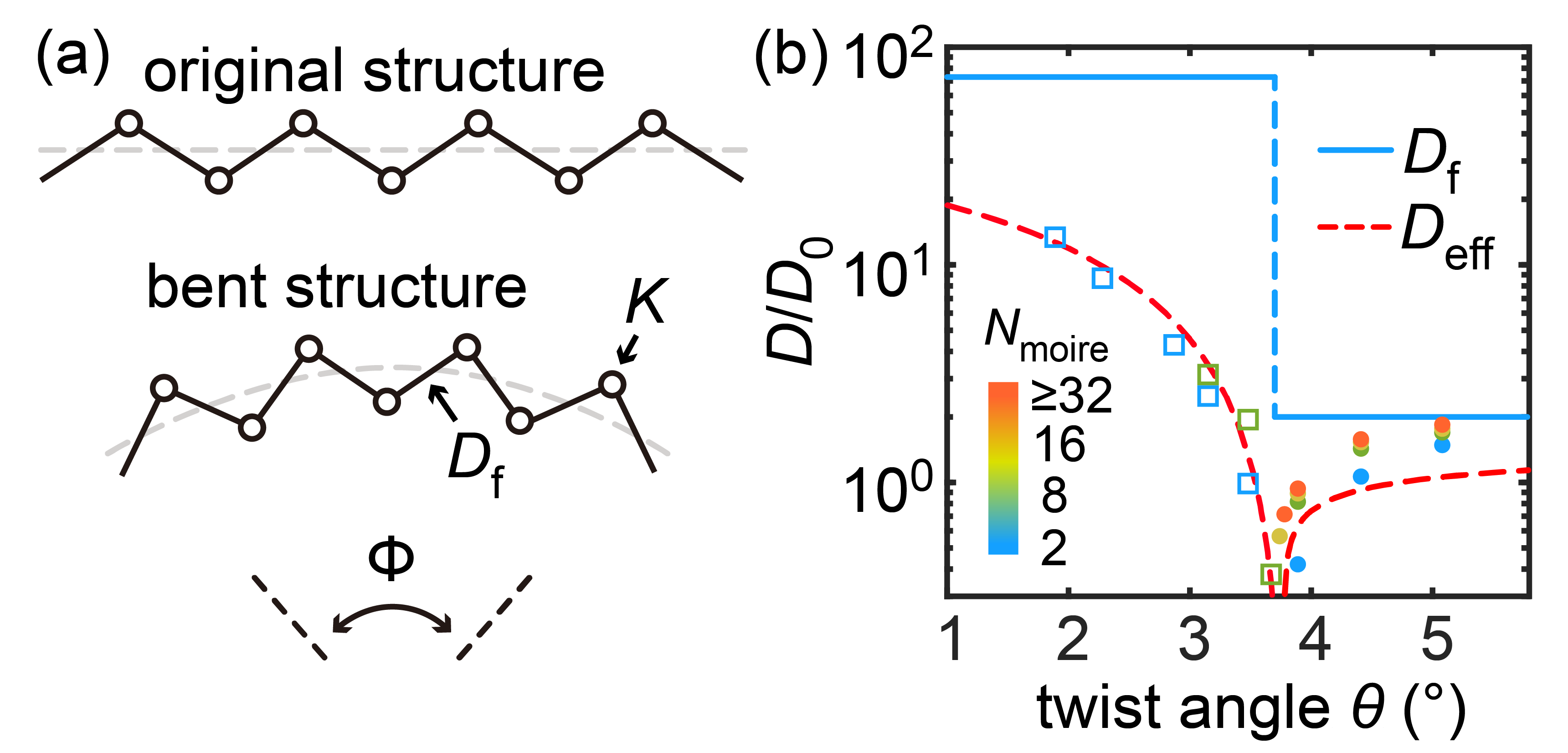}
\caption{Zigzag model of bending stiffness $D_\mathrm{eff}$ in a buckled bilayer. (a) Unbent and bent  models; (b) bending stiffness from real simulations of Fig.~\ref{fig:2} (circles and squares) and from zizgag model, Eq.~(3) (red dashed line). Assumed flat piece stiffness  $D_\mathrm{f}$ (blue line).
(more detail in SI. V).
}
\label{fig:3}
\end{figure}

Why does a {\it macroscopic} mechanical parameter like $D_x(\theta)$
drop critically at the {\it microscopic} buckling transition? We developed an analytical ``zigzag'' model that explains it.
As sketched in Fig.~\ref{fig:3}a the buckled structure can roughly be modeled as a  zigzag shape where flat (AB-commensurate) regions are separated by maximally bent (AA-centered) hinges. The total length of the system along the buckling direction is
$L_{x}=N_x l_x$.
Upon bending along $x$ both hinges and flats undergo deformation, and the free energy increase
with bending angle $\Phi$ 
is $ F(\Phi) = \frac{D_\mathrm{f} K l_y}{2N (2D_\mathrm{f} l_y + K l_x)}\Phi^2$
where $D_\mathrm{f}$ is the bending stiffness of the flat pieces, $K$ the angular stiffness of the hinges, and $l_y$ the bilayer size perpendicular to the bending direction $x$. Defining an effective bending stiffness
$F=\frac{D_\mathrm{eff} l_y}{2L_x}\Phi^2$, one obtains, using $l_y=\lambda$, 
\begin{equation}
    D_\mathrm{eff}= \frac{D_\mathrm{f}}{1+2D_\mathrm{f}/\sqrt{3}K}.
\end{equation}
We can now assume $D_\mathrm{f} = D_\mathrm{B} \approx 100~\mathrm{eV}$ of the flat pieces for $\theta<\theta_c$, dropping to $ D_\mathrm{f} = 2D_0=2.88~\mathrm{eV}$ for $\theta>\theta_c$ (Fig.~\ref{fig:3}b).
The collapse of $D_x(\theta)$ at $\theta_c$ is controlled by that of the hinge stiffness $K$, connected to the buckling frequency $\omega_\mathrm{\pm}$ by simple mechanics
\begin{equation}
K \propto \rho_\mathrm{2D} l_x^4 \omega_\mathrm{\pm}^2.
\end{equation}
Inserting $\omega_\mathrm{\pm}$ into Fig.~\ref{fig:3}b, theoretical and simulated bending stiffnesses agree fairly well, both in magnitude and in critical scaling (details in SI.V).
Thus the TBG bending stiffness collapse is a direct consequence of that of the buckling modes $\omega_\mathrm{\pm}$.
In return, the buckling criticality must be influenced by the bending one. The coexistence of these two coupled degrees of freedom, with important cross correlations, is likely to account for the unusual exponents.\\

We come next to two important properties predicted for the freestanding TBG buckled state, namely temperature behaviour, and the narrow band electronic structure at the magic twist angles.
\begin{figure}[ht!]
\centering
\includegraphics[width=\linewidth]{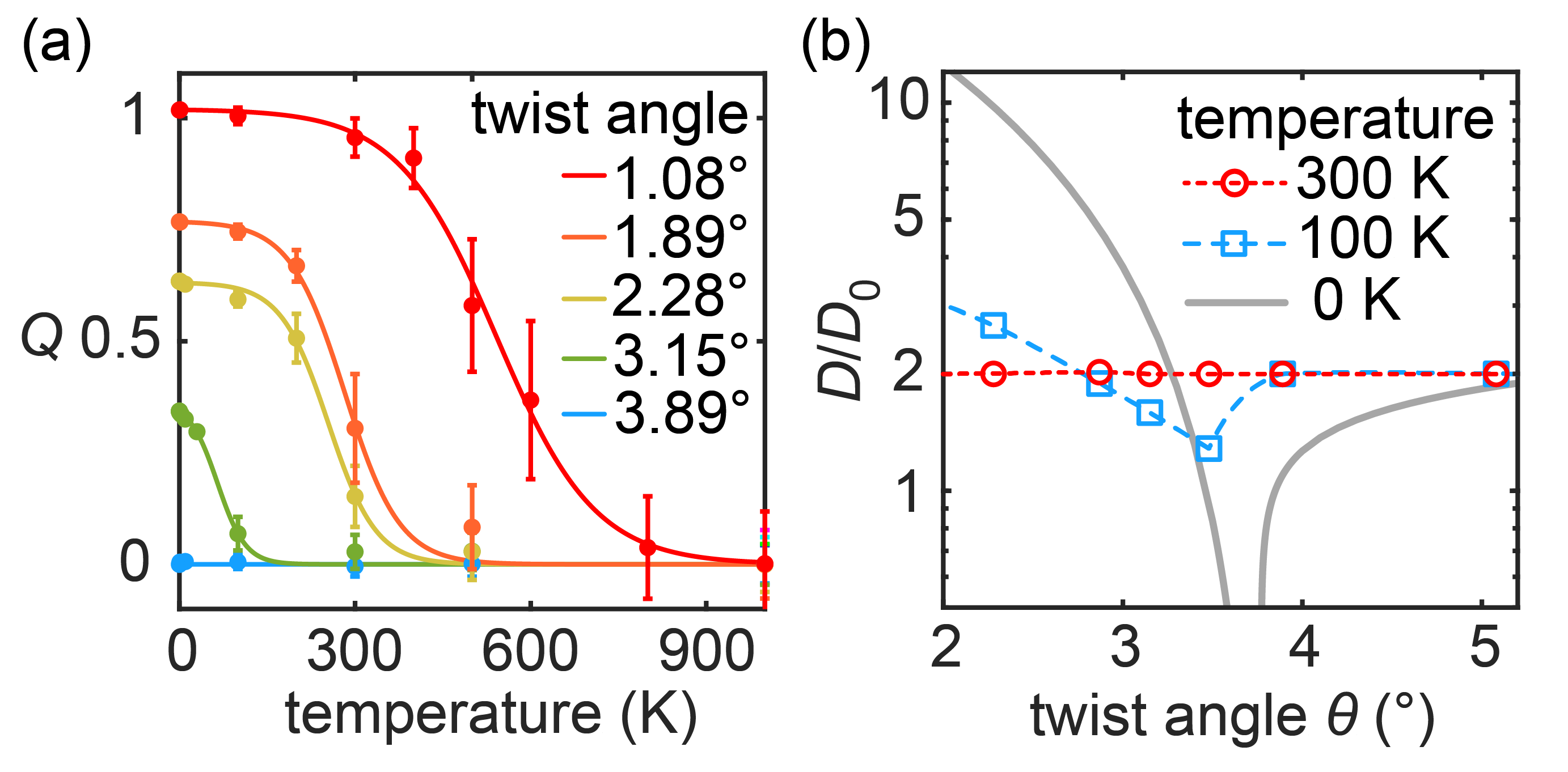}
\caption{Effect of temperature.
(a) Buckling order parameter at high temperatures. The expected critical behaviour
is smoothened by small simulation size (three $(2 \times 1)$ cells at each $\theta$).
(b) Bending stiffness at $T=$ 0 (gray, from Fig.~\ref{fig:2}), 100 K (blue) and 300 K (red). Note the extreme sensitivity to temperature. }
\label{fig:4}
\end{figure}

{\it Temperature}.
Finite temperature MD simulations show that at small twist angle the buckling persists up to high temperature. Flexural fluctuations, abundant and not gapped, do not cancel the buckling order parameter, which survives up to a remarkable $\approx$ 500 K at the magic twist angle (Fig.~\ref{fig:4}a). If bending could be ignored, this robust buckling order should drop at some $T_c$ with 3-state Potts universality, whose behaviour is critical as opposed to first order, despite  the presence of Landau 3rd order invariants \cite{Wu.RevModPhys.1982}
\bibnote{Under uniaxial stress or asymmetric boundary condition, such as sketched in Fig.~\ref{fig:1}a-b, that behavior might turn to Ising.}.
Unfortunately, size limitations obscure the high temperature behaviour, replacing it with the smooth crossovers of Fig.~\ref{fig:4}a (see SI. IV),  equally compatible with either continuous or discontinuous decays. 
Contrary to the buckling robustness, bending stiffness changes dramatically with temperature. The singularity near $\theta_c$ is quickly wiped out (Fig.~\ref{fig:4}b), the two layers eventually bending independently  despite the large buckling. Reflecting that, the flexural fluctuations of a TBG near $\theta_c$  should grow anomalously at low temperature, when $D_x$ is small, but not above.\\

\begin{figure}[ht!]
\centering
\includegraphics[width=\linewidth]{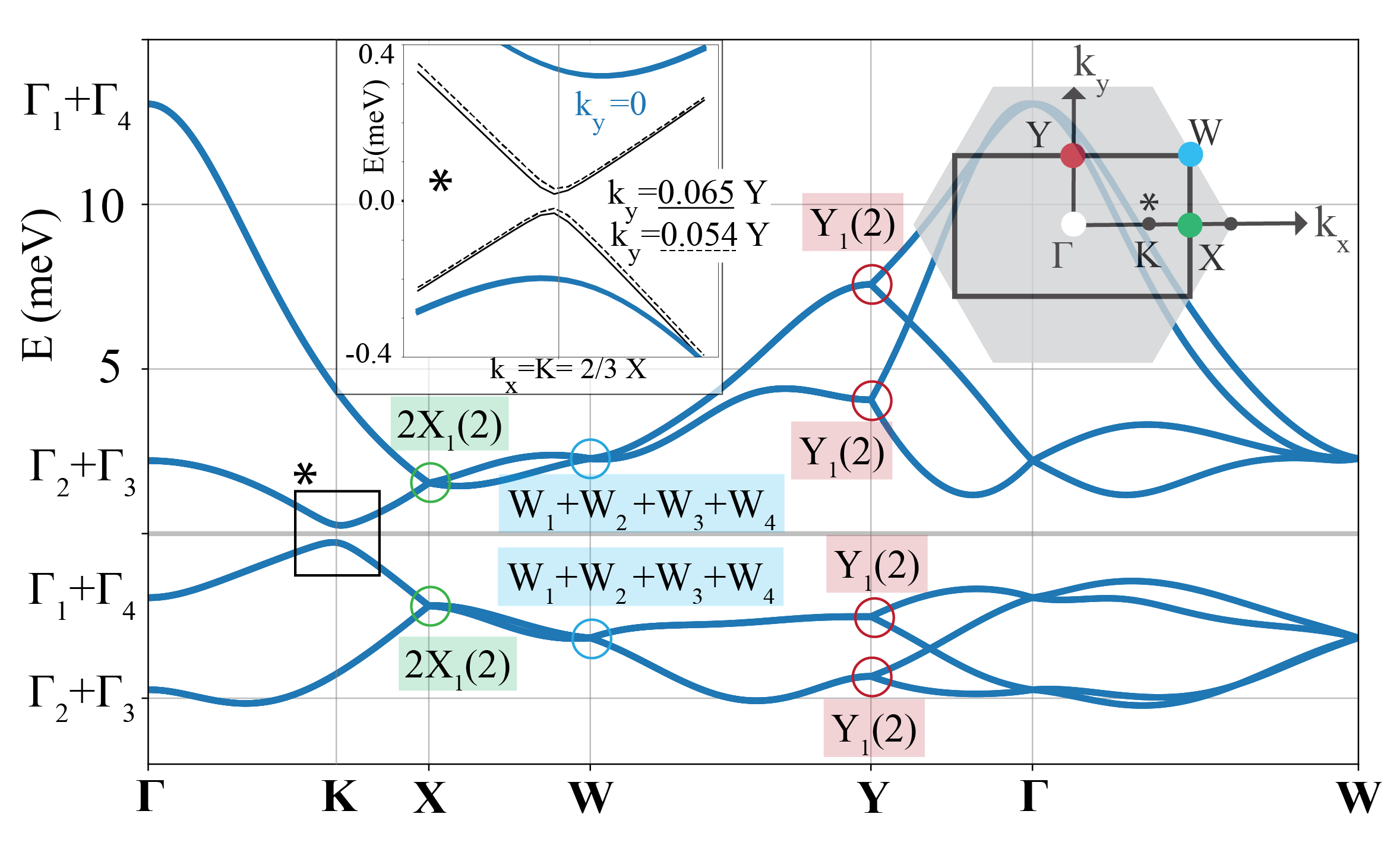}
\caption{The eight narrow bands of the $(2 \times 1)$ buckled TBG at the twist angle $\theta =1.08^{\circ}$ (see note [21]).
Inset $*$ shows how the gap closing point is now split into two, above and below (not shown) the point $K$ where it lies in the flat TBG.
}
\label{fig:5}
\end{figure}

{\it Electronic structure}. Near the magic twist angle $\theta_m =1.08^\circ$, the flat TBG has four ultra-narrow bands, whose physics has been at the center of much attention \cite{Bistritzer.PNAS.2011,Cao.nature.2018-1}.
What will happen of these bands in the freestanding TBG, where the buckling sets in? We should anticipate eight narrow bands, separated by zone boundary gaps caused by the large $(2 \times 1)$ distortion.
We carried out tight-binding calculations at $\theta_m$ and compared buckled and unbuckled TBG. Shown in Fig.~\ref{fig:5} \bibnote{ In order to be consistent with conventions of \cite{Cao.nature.2018-1} (Cao, Nature, 2018) the x and y directions in the electronic section are exchanged with respect to the structural part} are the 8 bands of the buckled state.
The bands, almost a factor 2 wider, display important novelties. First, unlike the unbuckled TBG (or even the non-optimal $(1 \times 1$) buckled state \cite{Rakib.commphys.2022}) the Dirac zero gap, formerly at point $K$, is now split into two close-by points $K \pm  k_y$ (see inset).
Second, and striking, zone boundary splittings at W and X points -- expected because $(2 \times 1$) buckling removes the $C_{3z}$ symmetry, and Bragg scattering should in principle take place -- do not occur.
This anomaly calls for a full symmetry analysis.

The buckled structure has, unlike the flat one, a non-symmorphic space group $P2_12_12$ (no. 18) that includes  $\big\{1\mid 0\big\}$, $\big\{2_{010}\mid 0,1/2,0\big\}$, 
$\big\{2_{100}\mid 1/2,0,0\big\}$ and $\big\{2_{001}\mid 1/2,1/2,0\big\}$, (in Seitz notation, with fractional translations) which we shortly denote as $E$, $C_{2y}$, $C_{2x}$ and $C_{2z}$, respectively.
The Bloch states of the 8 bands at the high symmetry points transform as the irreducible representations (irreps) of the corresponding little groups (see  Table~S3 in SI).
In our reference frame only $C_{2y}$ commutes with the generator of $U(1)$ valley symmetry, while $C_{2x}$, and thus $C_{2z}$, exchange the valley index -- as time-reversal symmetry $\hat T$ also does. It follows that, e.g., at the $\bd{\Gamma}$-point, $C_{2x}$ and valley $U(1)$ get promoted to $SU(2)$. That enforces degeneracy between eigenstates with opposite parity under $C_{2x}$ and same parity under $C_{2y}$ \cite{Mattia-PRX2019},
making $\Gamma_1$ degenerate with $\Gamma_4$, and $\Gamma_2$ with $\Gamma_3$.
The interplay between valley $U(1)$ and the non-symmorphic group's fractional translations has further consequences at zone boundary points $\bf{W}$ and $\bf{X}$.
Physical insight is obtained by switching to a single valley representation where valley index is conserved \cite{Bernevig-PRL2019}. This representation obeys the
magnetic space group $P2_12'_12'$ (no. 18.19) generated by $E$, $C_{2y}$, $\hat{T}\,C_{2x}$ and $\hat{T}\,C_{2z}$. 
The four flat bands per valley can be generated by Wannier orbitals centered at the Bernal stacked regions.
Their centers correspond to the Wyckoff positions $4c$ in the magnetic space group $P2_12'_12'$ \cite{Bilbao-1,*Bilbao-2,*Bilbao-3}, whose co-representation contains just the identity and thus allows a single irrep.
The single-valley elementary band representation is \cite{Bilbao-1,*Bilbao-2,*Bilbao-3}:
$ 2\Gamma_1(+)\oplus2\Gamma_2(-)$ and $2Y_1(+)\oplus2Y_2(-)$, where $(\pm)$ indicate the parity under $C_{2y}$;
$X_1X_2(2)$ and $W_1W_2(2)$, transforming under $C_{2y}$, respectively, as $\sigma_3$, the third Pauli matrix, and $i\sigma_3$ (see Table S3). Along the path $\mathbf{X}\to\mathbf{W}$, $k_y\in 0\to\pi$, the irreducible representation remains twofold degenerate and transforms under $C_{2y}$ like $\text{e}^{i k_y/2}\,\sigma_3$.
This now yields double degeneracies within a single valley. Bearing in mind that the two valleys must be further degenerate at all high-symmetry points and paths that are invariant under $C_{2x}$,
one  readily recovers the fourfold degeneracy at $\bf{W}$. In fact all ``accidental" degeneracies of the band structure in Fig.~\ref{fig:5} are explained in terms of single valley physics, enforced by non symmorphycity.

These one-electron degeneracies, and the ways they might be broken by interactions, represent an interesting question for freestanding TBG experiments, where topology should also play a role.  Since narrow bands do admit an elementary representation, the arguments used in the unbuckled case to diagnose a fragile topology \cite{Bernevig-PRL2019} do not strictly apply here.
The unexpected similarity of buckled bands to the $(2 \times 1)$ BZ folded unbuckled ones (Fig.~S12), nonetheless suggests that the topological properties remain similar. Thus \cite{Mattia-PRX2019,Blason-PRB2022} the interplay between Coulomb repulsion and electron-phonon coupling to Kekul\'e modes should split the degeneracies and open the gaps that are absent at the one electron level, stabilising topological insulators in $(2 \times 1)$ buckled TBG, giving rise to novel fractional fillings absent in the flat state (See  SI. IX).

In summary, several important phenomena are predicted to occur once freestanding TBG will be realized. 
First, a zigzag buckled state should set in with a critical behavior as a function of twist angle $\theta \to \theta_c \approx 3.7^\circ$.
At $\theta \approx 1^\circ$ the normals to the bilayer should deviate from $\hat{z}$ by a sizeable $\sim \pm 3^\circ$ (SI. VII), experimentally observable.
Second, the macroscopic bending stiffness, a crucial mechanical parameter for a membrane, should collapse at the buckling transition, giving rise to gigantic flexural fluctuations already at very low temperatures.
Third,  the buckling distortion should survive up to relatively high temperatures, whereas the bending stiffness anomaly will on the contrary dwindle upon heating.
Fourth, narrow electronic bands are predicted for the buckled magic TBG displaying unexpected single-valley degeneracies, to be broken by interactions, with the possibility of doubling the number of quantized fillings  upon gating. That should offer a richer playground for topological features and insulating states than for flat TBG. Other properties including kinetic and tribological behaviour will be addressed in follow-up work. Similar buckling phenomena could take place in freestanding bilayers of other 2D materials, now being pursued.\\

{\it Acknowledgments.} Work carried out under ERC ULTRADISS Contract No. 834402, with discussions with E. Meyer and M. Kisiel.
Support by the Italian Ministry of University and Research through PRIN UTFROM N. 20178PZCB5 is also acknowledged.
J.W. acknowledges the computing resources support from National Supercomputer Center in Tianjin.
M.F. acknowledges support from Italian Ministry of University and Research under the PRIN 2020 program, project No. 2020JLZ52N.

$\dagger$ these authors contributed equally.

\bibliographystyle{unsrt}
\bibliography{ref}

\begin{thebibliography}{10}

\bibitem{Geim.nature.2013}
A.~K. Geim and I.~V. Grigorieva.
\newblock {\em Nature}, 499(7459):419--425, 2013.

\bibitem{Neto.science.2016}
K.~S. Novoselov, A.~Mishchenko, A.~Carvalho, and A.~H.~Castro Neto.
\newblock {\em Science}, 353(6298):aac9439, 2016.

\bibitem{Yankowitz.NatRevPhys.2019}
M.~Yankowitz, Q.~Ma, P.~Jarillo-Herrero, and B.~J. LeRoy.
\newblock {\em Nat. Rev. Phys.}, 1(2):112--125, 2019.

\bibitem{Liu.NatRevMater.2016}
Y.~Liu, N.~O. Weiss, X.~Duan, H.~C. Cheng, Y.~Huang, and X.~Duan.
\newblock {\em Nat. Rev. Mater.}, 1(9):16042, 2016.

\bibitem{Bistritzer.PNAS.2011}
R.~Bistritzer and A.~H. MacDonald.
\newblock {\em Proc. Natl. Acad. Sci.}, 108(30):12233--12237, 2011.

\bibitem{Cao.nature.2018-1}
Y.~Cao, V.~Fatemi, A.~Demir, S.~Fang, S.~L. Tomarken, J.~Y. Luo, J.~D.
  Sanchez-Yamagishi, K.~Watanabe, T.~Taniguchi, E.~Kaxiras, R.~C. Ashoori, and
  P.~Jarillo-Herrero.
\newblock {\em Nature}, 556(7699):80--84, 2018.

\bibitem{Cao.nature.2018-2}
Y.~Cao, V.~Fatemi, S.~Fang, K.~Watanabe, T.~Taniguchi, E.~Kaxiras, and
  P.~Jarillo-Herrero.
\newblock {\em Nature}, 556(7699):43--50, 2018.

\bibitem{Butz.nature.2014}
B.~Butz, C.~Dolle, F.~Niekiel, K.~Weber, D.~Waldmann, H.~B. Weber, B.~Meyer,
  and E.~Spiecker.
\newblock {\em Nature}, 505(7484):533--537, 2014.

\bibitem{Ying.nc.2022}
Y.~Ying, Z.~Z. Zhang, J.~Moser, Z.~J. Su, X.~X. Song, and G.~P. Guo.
\newblock {\em Nat. Commun.}, 13(1):6392, 2022.

\bibitem{Dai.nanolett.2016}
S.~Dai, Y.~Xiang, and D.~J. Srolovitz.
\newblock {\em Nano Lett.}, 16(9):5923--5927, 2016.

\bibitem{Lamparski.2dmater.2020}
M.~Lamparski, B.~Van Troeye, and V.~Meunier.
\newblock {\em 2D Mater.}, 7(2):025050, 2020.

\bibitem{Rakib.commphys.2022}
T.~Rakib, P.~Pochet, E.~Ertekin, and H.~T. Johnson.
\newblock {\em Commun. Phys.}, 5(1):242, 2022.

\bibitem{Landau.elasticity}
L.~D. Landau, E.~M. Lifshitz, A.~M. Kosevich, and L.~P. Pitaevskii.
\newblock volume 7 {\it Theory of Elasticity}.
\newblock Elsevier, 1986.

\bibitem{Lu.JphysD.2009}
Q.~Lu, M.~Arroyo, and R.~Huang.
\newblock {\em J. Phys. D}, 42(10):102002, 2009.

\bibitem{Han.natmat.2020}
E.~Han, J.~Yu, E.~Annevelink, J.~Son, D.~A. Kang, K.~Watanabe, T.~Taniguchi,
  E.~Ertekin, P.~Y. Huang, and A.~M. van~der Zande.
\newblock {\em Nat. Mater.}, 19(3):305--309, 2020.

\bibitem{Yu.AdvMat.2021}
J.~Yu, E.~Han, M.~A. Hossain, K.~Watanabe, T.~Taniguchi, E.~Ertekin, A.~M.
  van~der Zande, and P.~Y. Huang.
\newblock {\em Adv. Mater.}, 33(9):2007269, 2021.

\bibitem{Lee.nanolett.2012}
J.~U. Lee, D.~Yoon, and H.~Cheong.
\newblock {\em Nano Lett.}, 12(9):4444--4448, 2012.

\bibitem{Wang.prl.2019}
G.~Wang, Z.~Dai, J.~Xiao, S.~Feng, C.~Weng, L.~Liu, Z.~Xu, R.~Huang, and
  Z.~Zhang.
\newblock {\em Phys. Rev. Lett.}, 123:116101, 2019.

\bibitem{Wu.RevModPhys.1982}
F.~Y. Wu.
\newblock {\em Rev. Mod. Phys.}, 54:235--268, 1982.

\bibitem{Note1}
Under uniaxial stress or asymmetric boundary condition, such as sketched in
  Fig.~\ref {fig:1}a-b, that behavior might turn to Ising.

\bibitem{Note2}
In order to be consistent with conventions of \cite {Cao.nature.2018-1} (Cao,
  Nature, 2018) the x and y directions in the electronic section are exchanged
  with respect to the structural part.

\bibitem{Mattia-PRX2019}
M.~Angeli, E.~Tosatti, and M.~Fabrizio.
\newblock {\em Phys. Rev. X}, 9:041010, 2019.

\bibitem{Bernevig-PRL2019}
Z.~Song, Z.~Wang, W.~Shi, G.~Li, C.~Fang, and B.~A. Bernevig.
\newblock {\em Phys. Rev. Lett.}, 123:036401, 2019.

\bibitem{Bilbao-1}
M.~I. Aroyo, J.M. Perez-Mato, D.~Orobengoa, E.~Tasci, G.~De~La~Flor, and
  A.~Kirov.
\newblock {\em Bulg. Chem. Commun.}, 43(2):183 – 197, 2011.

\bibitem{Bilbao-2}
M.~I. Aroyo, J.~M. Perez-Mato, C.~Capillas, E.~Kroumova, S.~Ivantchev,
  G.~Madariaga, A.~Kirov, and H.~Wondratschek.
\newblock {\em Z. Kristallogr. Cryst. Mater.}, 221(1):15--27, 2006.

\bibitem{Bilbao-3}
M.~I. Aroyo, A.~Kirov, C.~Capillas, J.~M. Perez-Mato, and H.~Wondratschek.
\newblock {\em Acta Crystallogr. A}, 62(2):115--128, 2006.

\bibitem{Blason-PRB2022}
A.~Blason and M.~Fabrizio.
\newblock {\em Phys. Rev. B}, 106:235112, 2022.

\end{thebibliography}


\begin{thebibliography}{10}

\bibitem{Trambly.nanolett.2010}
G.~Trambly~de Laissardière, D.~Mayou, and L.~Magaud.
\newblock {\em Nano Lett.}, 10(3):804--808, 2010.

\bibitem{Kolmogorov.prb.2005}
A.~N. Kolmogorov and V.~H. Crespi.
\newblock {\em Phys. Rev. B}, 71:235415, 2005.

\bibitem{Ouyang.nanolett.2018}
W.~Ouyang, D.~Mandelli, M.~Urbakh, and O.~Hod.
\newblock {\em Nano Lett.}, 18(9):6009--6016, 2018.

\bibitem{Brenner.jpcm.2002}
D.~W. Brenner, O.~A. Shenderova, J.~A. Harrison, S.~J. Stuart, B.~Ni, and S.~B.
  Sinnott.
\newblock {\em J. Phys. Condens.}, 14(4):783--802, 2002.

\bibitem{Plimpton.jcp.1995}
S.~Plimpton.
\newblock {\em J. Comput. Phys.}, 117(1):1--19, 1995.

\bibitem{Thompson.compphyscomm.2022}
A.~P. Thompson, H.~M. Aktulga, R.~Berger, D.~S. Bolintineanu, W.~M. Brown,
  P.~S. Crozier, P.~J. {in 't Veld}, A.~Kohlmeyer, S.~G. Moore, T.~D. Nguyen,
  R.~S., M.~J. Stevens, J.~Tranchida, C.~Trott, and S.~J. Plimpton.
\newblock {\em Comput. Phys. Commun.}, 271:108171, 2022.

\bibitem{Bitzek.prl.2006}
E.~Bitzek, P.~Koskinen, F.~G\"ahler, M.~Moseler, and P.~Gumbsch.
\newblock {\em Phys. Rev. Lett.}, 97:170201, 2006.

\bibitem{Han.natmat.2020}
E.~Han, J.~Yu, E.~Annevelink, J.~Son, D.~A. Kang, K.~Watanabe, T.~Taniguchi,
  E.~Ertekin, P.~Y. Huang, and A.~M. van~der Zande.
\newblock {\em Nat. Mater.}, 19(3):305--309, 2020.

\bibitem{Yu.AdvMat.2021}
J.~Yu, E.~Han, M.~A. Hossain, K.~Watanabe, T.~Taniguchi, E.~Ertekin, A.~M.
  van~der Zande, and P.~Y. Huang.
\newblock {\em Adv. Mater.}, 33(9):2007269, 2021.

\bibitem{Martyna.jcp.1994}
G.~J. Martyna, D.~J. Tobias, and M.~L. Klein.
\newblock {\em J. Chem. Phys.}, 101(5):4177--4189, 1994.

\bibitem{Fasolino.nature.2007}
A.~Fasolino, J.~H. Los, and M.~I. Katsnelson.
\newblock {\em Nat. Mater.}, 6(11):858--861, 2007.

\bibitem{Thomas.prb.2010}
J.~A. Thomas, J.~E. Turney, R.~M. Iutzi, C.~H. Amon, and A.~J.~H. McGaughey.
\newblock {\em Phys. Rev. B}, 81:081411, 2010.

\bibitem{slater1954LCAO}
J.~C. Slater and G.~F. Koster.
\newblock {\em Phys. Rev.}, 94:1498--1524, 1954.

\bibitem{Mattia-PRX2019}
M.~Angeli, E.~Tosatti, and M.~Fabrizio.
\newblock {\em Phys. Rev. X}, 9:041010, 2019.

\bibitem{Bernevig-PRL2019}
Z.~Song, Z.~Wang, W.~Shi, G.~Li, C.~Fang, and B.~A. Bernevig.
\newblock {\em Phys. Rev. Lett.}, 123:036401, 2019.

\end{thebibliography}

\end{document}

%\preprint{APS/123-QED}

\title{Supplementary Information \\
\textbf{Bending Stiffness Collapse, Buckling, Topological Bands of Freestanding Twisted Bilayer Graphene}} %new title

\author{Jin Wang}
\affiliation{International School for Advanced Studies (SISSA), I-34136 Trieste, Italy}
\author{Ali Khosravi}
\affiliation{International School for Advanced Studies (SISSA), I-34136 Trieste, Italy}
\affiliation{International Centre for Theoretical Physics, I-34151 Trieste, Italy}
\affiliation{CNR-IOM, Consiglio Nazionale delle Ricerche - Istituto Officina dei Materiali, c/o SISSA,  34136, Trieste, Italy}
\author{Andrea Silva}
\affiliation{CNR-IOM, Consiglio Nazionale delle Ricerche - Istituto Officina dei Materiali, c/o SISSA,  34136, Trieste, Italy}
\affiliation{International School for Advanced Studies (SISSA), I-34136 Trieste, Italy}

\author{Michele Fabrizio}
\affiliation{International School for Advanced Studies (SISSA), I-34136 Trieste, Italy}

\author{Andrea Vanossi}
\affiliation{CNR-IOM, Consiglio Nazionale delle Ricerche - Istituto Officina dei Materiali, c/o SISSA,  34136, Trieste, Italy}
\affiliation{International School for Advanced Studies (SISSA), I-34136 Trieste, Italy}

\author{Erio Tosatti}
\email{tosatti@sissa.it}
\affiliation{International School for Advanced Studies (SISSA), I-34136 Trieste, Italy}
\affiliation{International Centre for Theoretical Physics, I-34151 Trieste, Italy}
\affiliation{CNR-IOM, Consiglio Nazionale delle Ricerche - Istituto Officina dei Materiali, c/o SISSA,  34136, Trieste, Italy}

\date{\today}

\maketitle

\tableofcontents
\newpage

\section{Simulation details}

Twisted bilayer graphene (TBG) with periodic boundary conditions (PBC) along $x$ and $y$ directions are constructed for a discrete set of twist angles $\theta$ ranging from $1.08^\circ$ to $30^\circ$  chosen for supercells of reasonable sizes \cite{Trambly.nanolett.2010}.
For the freestanding systems, no additional constraints are imposed along the out-of-plane (i.e., $z$) direction. To describe supported systems, $z$-direction springs with spring constant $k_z$ are tethered to each carbon atoms to mimic the constraints from the substrate. The interlayer and intralayer interaction is described by the registry-dependent Kolmogorov-Crespi potential with local normals \cite{Kolmogorov.prb.2005,Ouyang.nanolett.2018} and REBO force field respectively \cite{Brenner.jpcm.2002}.
All simulations are performed with open-source code LAMMPS \cite{Plimpton.jcp.1995,Thompson.compphyscomm.2022}.

{\it Structural optimization.}
During the optimization,  the simulation box adaptively changes size, so that the in-plane stress is fixed to zero, $p_{xx}=p_{yy}=0$. 
The FIRE  algorithm \cite{Bitzek.prl.2006} is used to minimize energy during structural optimization (together with conjugate gradient algorithms with several loops to optimize the box size).
Minimization stopped when the largest single atom force $|F_i| < 10^{-6}~\mathrm{eV/\AA}$. Unless otherwise specified, all structural optimization used this convergence criterion. 

{\it Bending simulation.}
A compression protocol \cite{Han.natmat.2020,Yu.AdvMat.2021} is used to extract the bending stiffness of TBG. As shown in Fig.~2 in the main text, by compressing the simulation box (along $x$ direction), the out-of-plane corrugation appears with height $h$ (in bending simulations, $k_z=0$).
The compression strain $\varepsilon=(L_0-L)/L_0$ in our bending simulations is chosen to be $\varepsilon \lesssim 0.1 \%$, thus the ratio between the bending corrugation height ($h\sim L_0 \sqrt{\varepsilon }/\pi$) and the length $h/L_0 \ll 1$.
Since the free energy of the system increases during the compression-induced bending, the system will spontaneously oscillate after the release of the boundary constraints. From  that we obtain the flexural oscillation frequency and thus the bending stiffness. In order to achieve free oscillations, the Nos\'e-Hoover thermostat and barostat are applied to the whole system during the simulation \cite{Martyna.jcp.1994} with $T \to 0$~K and $p_{xx}=p_{yy}=0$. The flexural frequencies are checked to be independent of the damping coefficient used.

\newpage
\section{Buckling order parameter at variable twist angle}

As in main text, we start at finite temperature with the same  order parameter definition $Q^*$ for the $(2 \times 1)$ buckling distortion in the form:
\begin{equation}
        Q^* =\frac{a_\mathrm{Gr}}{2 L_x L_y} \left \langle \left| \sum\limits_{n=1}^{N} z_n \exp\left (-\frac{2 \pi i}{l_x} x_n\right ) \right| \right \rangle 
\label{eq:S1}
\end{equation}
where $L_x=N_x l_x$ and $L_y=N_y l_y$ is the size of the %TBG ,
box,
$l_x\times l_y$ defines the size of one $(2\times 1)$ moir\'e cell, $N_x$ and $N_y$ are the number of replicas along $x$ and $y$ directions.
For the smallest simulation box ($N_x=N_y=1$) and at $T=0$ where the modulus is redundant, Eq.~(\ref{eq:S1}) coincides with  Fig.1 in maintext.
At nonzero temperatures, two problems arise with this definition of order parameter, a quantity that should represent the Bragg scattering magnitude produced by the symmetry breaking buckling distortion and nothing else.
The first problem is that the coordinate difference of two separate TBG layer centers-of-mass exhibits a noticeable $(x, y)$ random walk,
permitted by the superlubric nature of their incommensurate contact, and by the finite supercell size. The larger the size, and the lower the temperature, the smaller this artifact.
Because it gives rise to complex values of $Q$, we can eliminate it by taking the complex modulus as is done in $Q^*$. \\

The second problem is that temperature causes fluctuations with $(2 \times 1)$ periodicities, along with all other wavelengths, even in absence of distortions. This unwanted additional term had better be subtracted for the order parameter to yield correctly the  distortion magnitude, representing the $(2 \times 1)$ Bragg reflection magnitude of a hypothetical scatterer. This correction will be discussed and eliminated in subsequent section III.\\

\begin{figure}[ht!]
\centering
\includegraphics[width=0.75\linewidth]{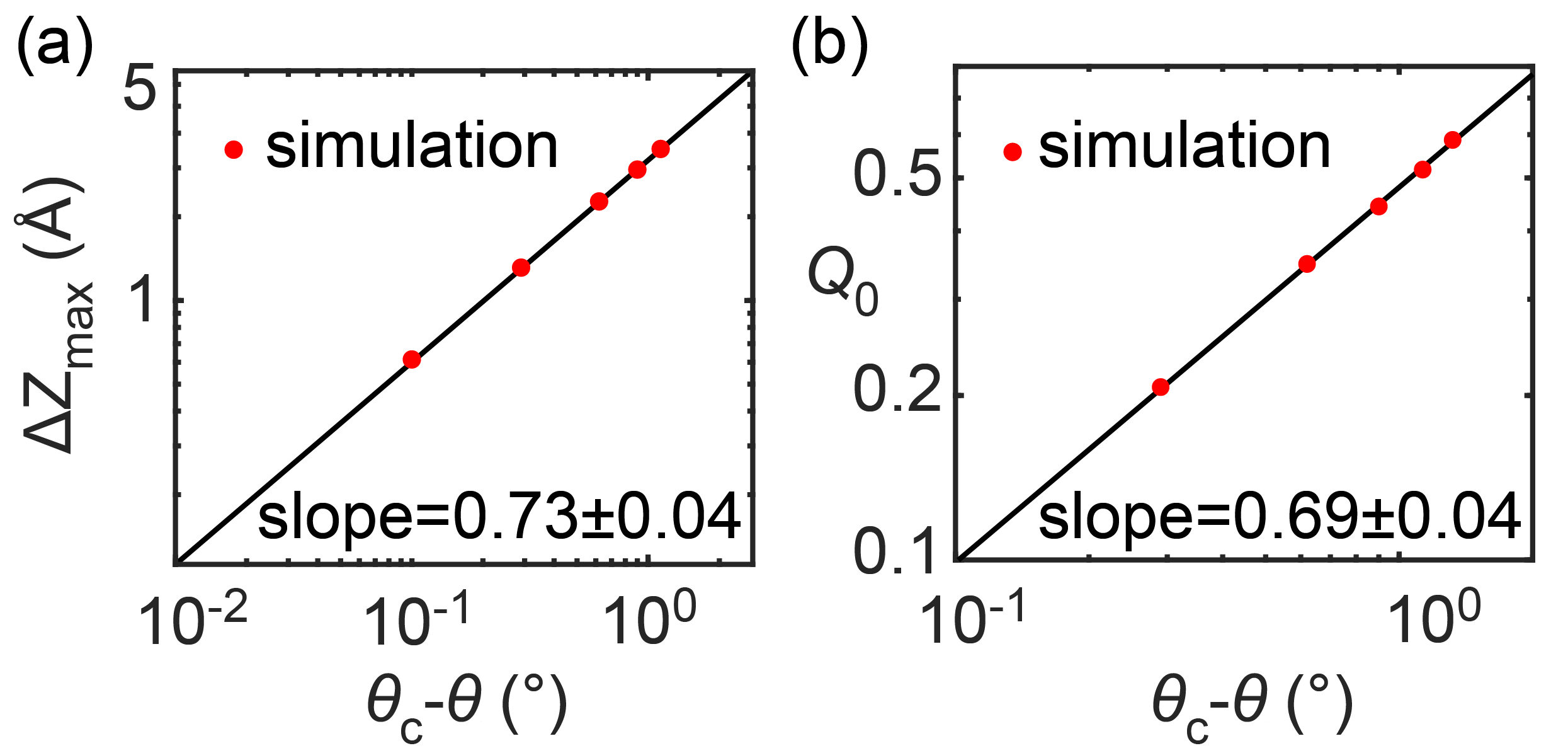}
\caption{Fitting of the out-of-plane corrugation $\Delta z_\mathrm{max}$ and order parameters $Q_0$ of the buckling systems as a function of twist angle $\theta$, at $T=0$~K. The critical angle is $\theta_c=3.77^\circ$. Parameters used here are $T=0$~K, $N_x=3$ and $N_y=1$.
}
\label{fig:S1}
\end{figure}

From its definition, it is apparent that there is a positive correlation between the order parameter $Q_0$ and the maximum out-of-plane deformation $\Delta z_\mathrm{max}=\max{(z)}-\min{(z)}$. For systems at $T=0$~K, we could use either parameter to represent the out-of-plane buckling magnitude. %intensity''. 
As shown in Fig.~\ref{fig:S1}, for buckled structures ($\theta<\theta_c$), both $\Delta z_\mathrm{max}$ and $Q_0$ are well fit by power laws, with an exponent $\approx 0.7(0)$. For $\theta>\theta_c$, the order parameters is zero -- no buckling.

\clearpage
\section{Temperature corrected order parameter}
As mentioned above, the order parameter definition needs to be corrected in the case of finite temperature.

A straight thermal fluctuation term $R(T)$ with buckling-unrelated $(2 \times 1)$ periodicity is to be subtracted from  $Q^*$:
\begin{equation}
\begin{aligned}
    Q=Q^*-R
\end{aligned}
\label{eq:S2}
\end{equation}
where $R$ should be the nonzero $(2 \times 1)$ Fourier amplitude due to ordinary thermal fluctuations, if the TBG remained
hypothetically unbuckled. As an approximation to that, we extrapolate to $T \leq T_c$ the $(2 \times 1)$ Fourier amplitude evaluated at $T \gtrsim T_c$, where the distortion has disappeared
\begin{equation}
    R (T \lesssim T_c)  = Q^* (T \gtrsim T_c)
\label{eq:S3}
\end{equation}
This thermal fluctuation with $k=2\pi/l_x$
further depends on the size of the bilayer and of course on temperature \cite{Fasolino.nature.2007},
$R(T) \propto l_x \sqrt{T}$, so that finally
\begin{equation}
    R(T)\propto \theta^{-1} \sqrt{T}
\label{eq:S5}
\end{equation}

\begin{figure*}[ht!]
\centering
\includegraphics[width=0.8\linewidth]{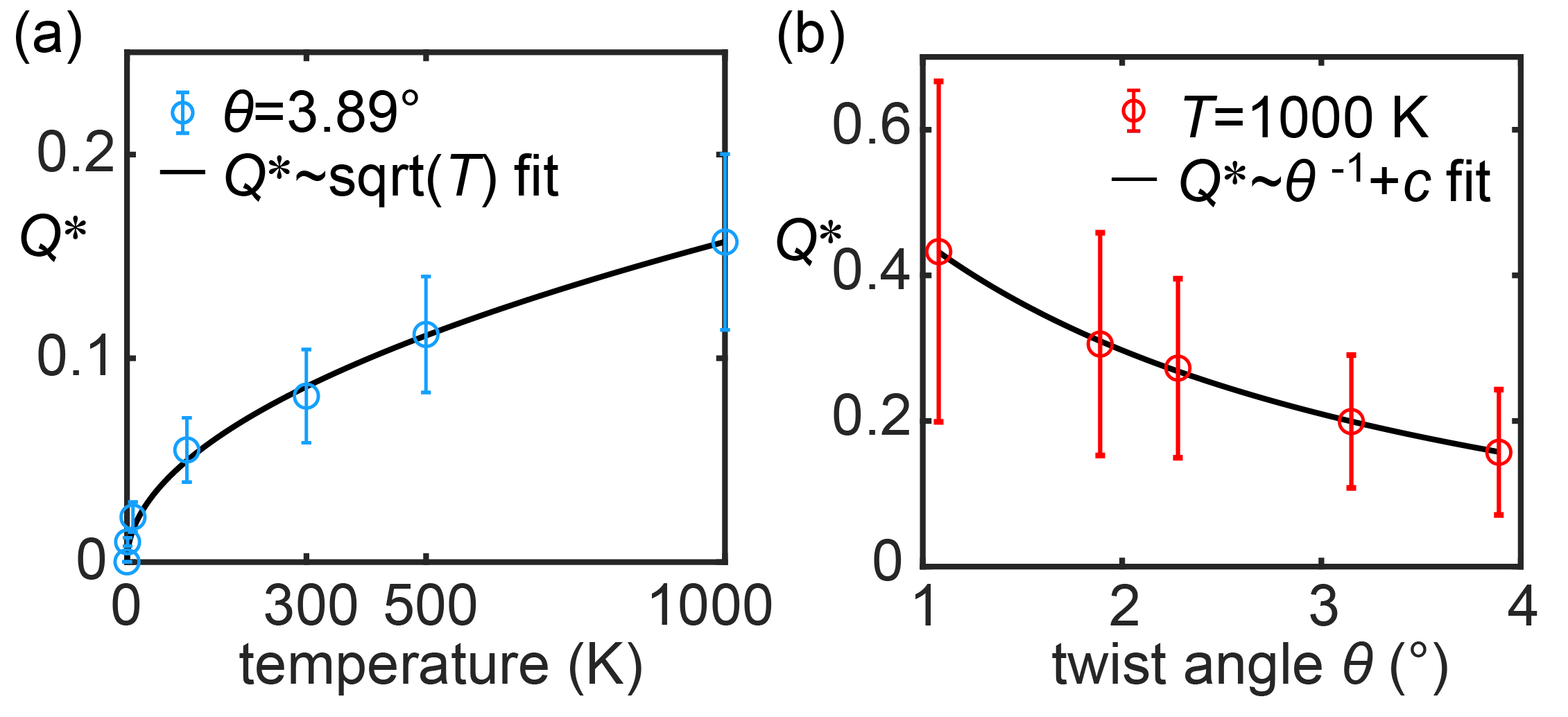}
\caption{Verification of Eq.~\eqref{eq:S5}. (a) The simulated order parameter $Q^*$ scales as $T^{1/2}$. (b) The simulated order parameter $Q^*$ scales as $\theta^{-1}$. Except for temperature, parameter used here are same as those in Fig.~\ref{fig:S1}.}
\label{fig:S2}
\end{figure*}

The thermal $(2 \times 1)$ background $R$ is extracted at $\theta=3.89^\circ$,
and its temperature dependence, needed for the extrapolation, is further verified at $T>T_c$, where buckling  is absent. 
As shown in Fig.~\ref{fig:S2}, the scaling given by Eq.~\eqref{eq:S5} is reasonable. 

\begin{figure}[ht!]
\centering
\includegraphics[width=0.8\linewidth]{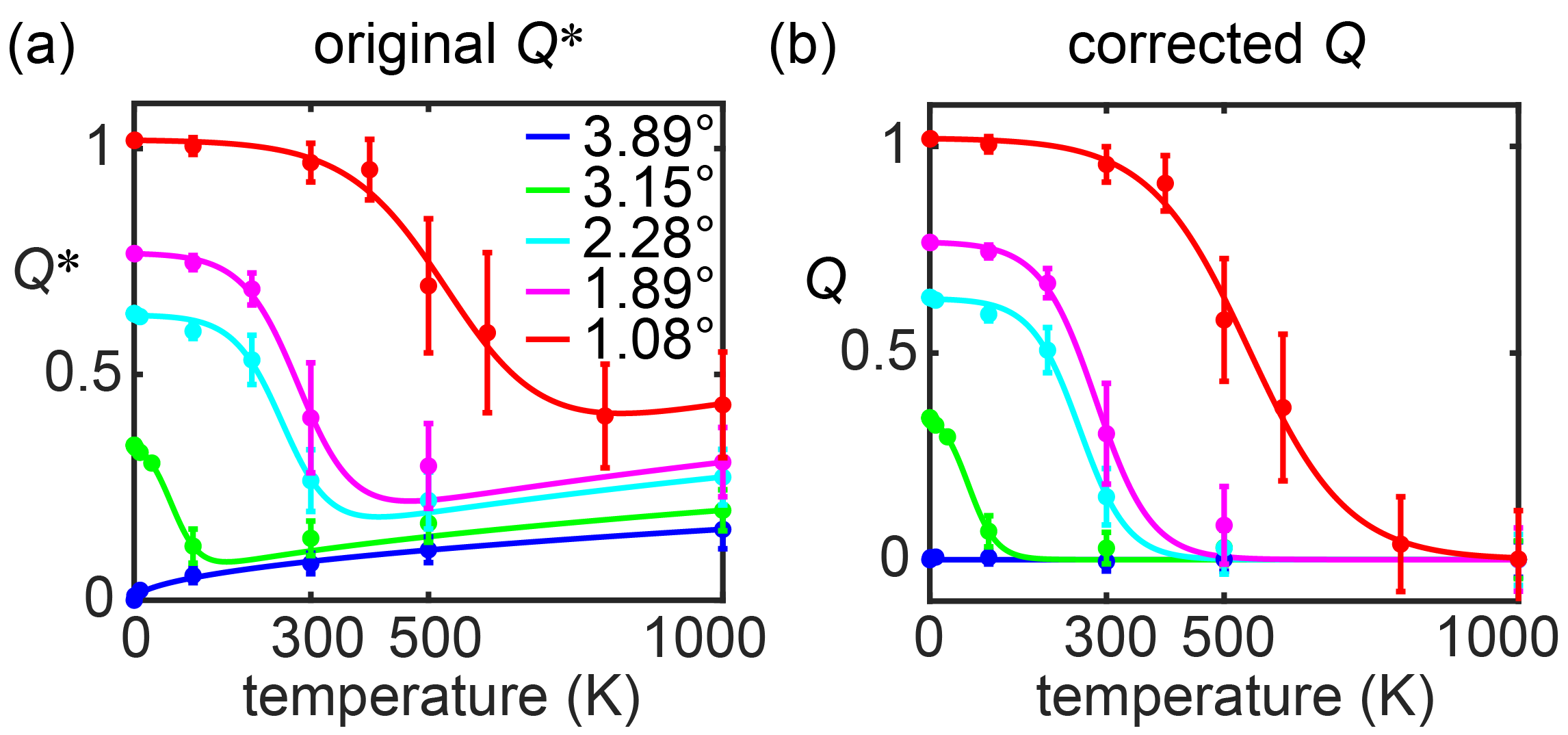}
\caption{(a) Uncorrected order parameter $Q^*$
for different twist angles. The solid line fit is based on a smoothed Heaviside step function (see text). (b) Corrected order parameter $Q$ (thermal background term subtracted).
Parameters used here are $N_x=3$ and $N_y=1$.
}
\label{fig:S3}
\end{figure}

A direct comparison of the original $Q^*$ and the correct $Q$ is shown in Fig.~\ref{fig:S3}. 

\newpage
\section{Unbuckling temperature $T_c$}

With the corrected parameter $Q$, we can estimate the unbuckling temperature $T_c$. First, it is generally seen that flexural fluctuations, large already at relatively low temperatures, play very little role and the buckling order parameter $Q$ remains close to $Q_0$ so long as ($T \ll T_c$) . At very high temperatures  ($T \gg T_c$) conversely, buckling disappears, and $Q$ =0.

The actual transition between the two regimes, ideally sharp at infinite size, (and presumably with order parameter exponent $\beta$ either  1/8  if Ising, or 1/9 if 3-state Potts) is artificially smoothed into a soft crossover at our small simulation sizes. A size-smoothed Heaviside step function (approximated as a sigmoid)
$S(T-T_c)$ is used to crudely account for this behaviour
$Q(T)=Q_t S$,
where $Q_t=Q_t(T)$ is the (unknown) true order parameter
and  $S(T-T_c) = \{1+\exp[(T-T_c)/T_f]\}^{-1}$, with $T_f$ a parameter characterizing the finite-size smoothing.
We found that a good fit of the finite size crossover of Fig.~\ref{fig:S3}b can already be obtained by crudely setting $Q_t(T) = Q_0$ (a choice compatible with a first order transition, but of course not proving it), with the transition temperature $T_c$ as the only parameter.
Values of $T_c$ for different twist angles are shown, along with the fitting parameter $T_f$ in Table S1.

\begin{table}[t]
\caption{\label{tab:table1} Estimated unbuckling temperature 
$T_c$ and finite size fitting parameter $T_f$ for different twist angles.}
\begin{ruledtabular}
\begin{tabular}{cccccc}
twist angle $\theta$ ($^\circ$) & 1.08 & 1.89 & 2.28 & 3.15 & 3.89\\
\hline
$T_c$ (K) & 541 & 283 & 254 & 65.2 & 0 \\
$T_f$ (K) & 83.8 & 46.9 & 40.8 & 24.0 & $\setminus$ \\
\end{tabular}
\end{ruledtabular}
\end{table}

As shown in Fig.~\ref{fig:S3}b, fits based on this crude trial (solid lines) agree well with the MD simulations (color-matched points).
The unbuckling transition temperature $T_c$ so obtained for  magic twist TBG (see Fig.~\ref{fig:S3}b and Table~\ref{tab:table1}) is as high as $\approx 500$~K -- one should by all means be able to observe this buckled structure at room temperature.

\clearpage
\section{Zigzag model details}
Here we give details about the zigzag model and discuss its applicability to understand the bending stiffness of TBG and the scaling exponents.
\subsection{Effective bending stiffness}

\begin{figure}[ht!]
\centering
\includegraphics[width=0.5\linewidth]{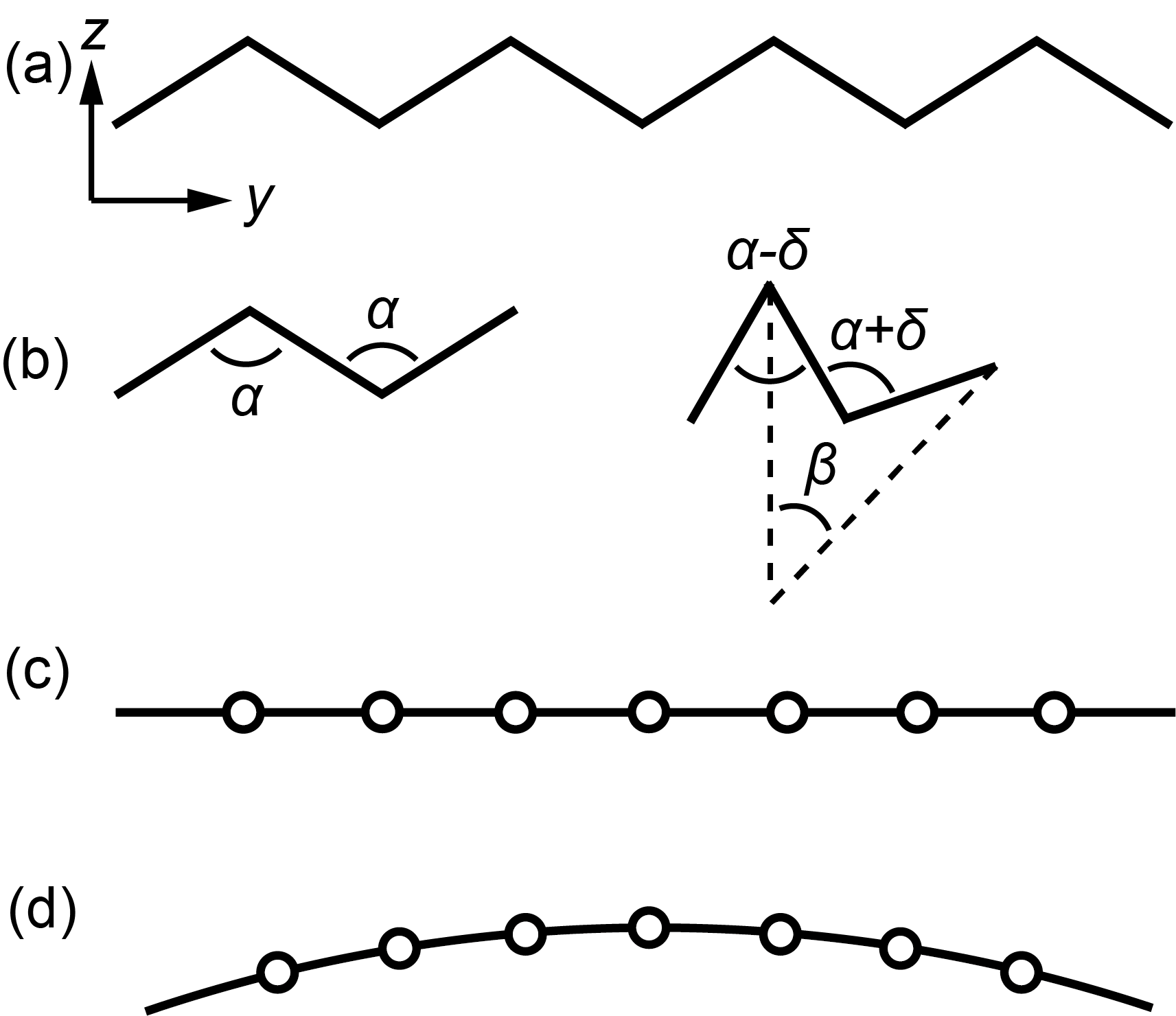}
\caption{Schematic diagram of the hinge model for the buckling structure.}
\label{fig:S4}
\end{figure}

The model is illustrated in Fig.~\ref{fig:S4}a -- a simplified zigzag out-of-plane mechanical structure. When this model structure is bent (Fig.~\ref{fig:S4}b), the hinge angles deform. The model (Fig.~\ref{fig:S4}c)
consists of flat (AB-like) regions and hinge (AA-like) regions. The total length of the system (along the bending direction) is $L_{x}$, where $L_{x}=N_x l_x$.

When the model is bent (Fig.~\ref{fig:S4}d), both hinge and the flat regions undergo the bending deformations, the relative free energy increase of the system is
\begin{equation}
    F (\Phi)=\frac{D_\mathrm{f} K l_x l_y \Phi^2}{2L_x(2D_\mathrm{f} l_y+ K l_x)}
    \label{eq:S7}
\end{equation}
where $D_\mathrm{f}$ is the bending stiffness of the flat region, $K$ is the angular stiffness of hinges, and $\Phi$ is the total bending angle. Compared to the bending energy expressed by the effective bending stiffness of the whole system,
\begin{equation}
    E=\frac{D_\mathrm{eff} l_y \Phi^2}{2 L_x}
    \label{eq:S8}
\end{equation}
one gets the effective bending stiffness for the zigzag buckled model:
\begin{equation}
    D_\mathrm{eff}=\frac{D_\mathrm{f}}{1+2D_\mathrm{f}/\sqrt{3}K}
    \label{eq:S9}
\end{equation}
where we used $l_x=\sqrt{3}\lambda$ and $l_y=\lambda$. The bending stiffness of the flat region $D_\mathrm{f}$ is approximated by:
\begin{equation}
    D_\mathrm{f}=\left\{
    \begin{aligned}
    D_\mathrm{B} \approx 100~\mathrm{eV} &, & (\theta<\theta_c)\\
    2 D_0=2.88~\mathrm{eV} &, & (\theta>\theta_c)
    \end{aligned}
    \right.
    \label{eq:S10}
\end{equation}
To estimate the bending stiffness from Eq.~(\ref{eq:S9}), the only undetermined parameter is the hinge stiffness $K$.

\subsection{Hinge stiffness K}

Here we provide simulation details that reveal the magnitude and $\theta$-dependence of the hinge stiffness $K$.
Since the stiffness is proportional to the square of buckling frequency, $K \propto  \rho_{\rm 2D} l^4 \omega_{\pm}^2$, one can firstly extract the $\theta$-dependence of the buckling frequency $\omega(\theta)$.

\begin{figure}[ht!]
\centering
\includegraphics[width=0.8\linewidth]{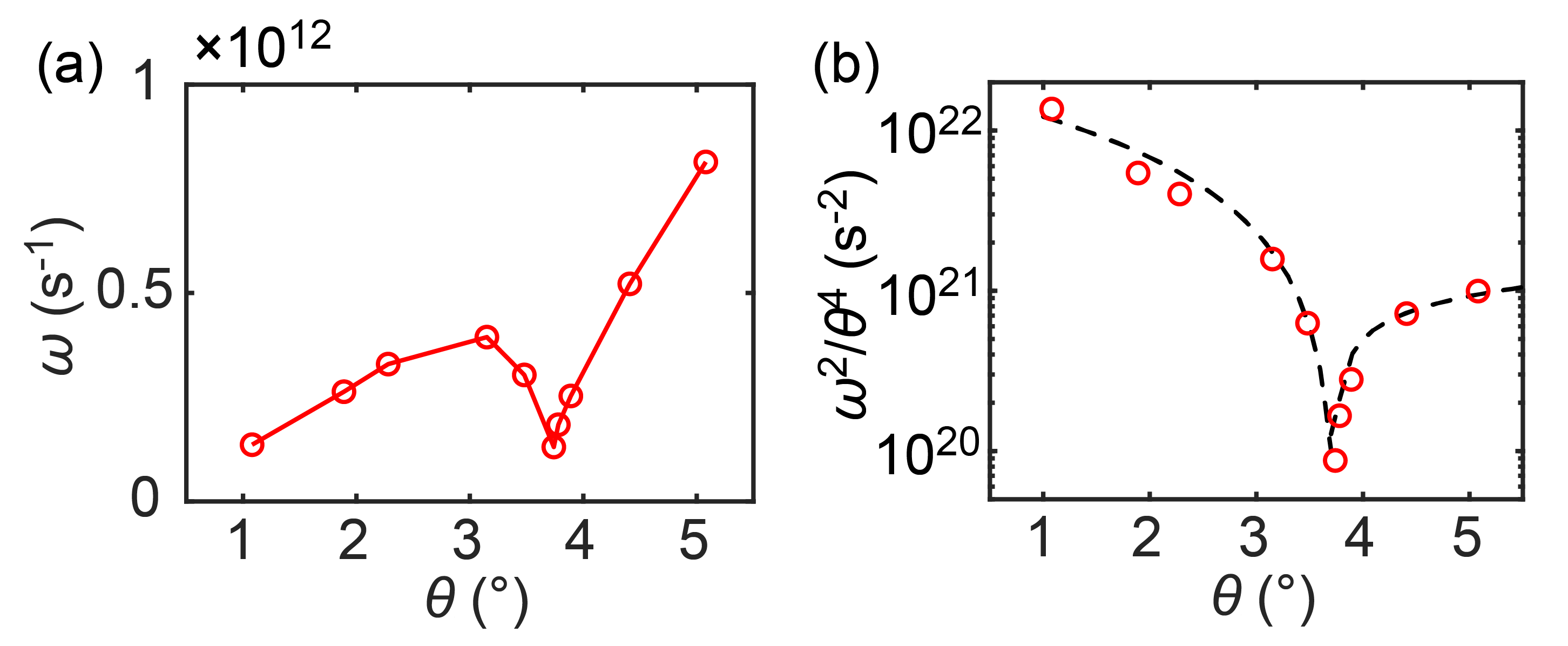}
\caption{(a) Dependence of the buckling frequency $\omega$ upon the twist angle $\theta$.
(b)Twist angle dependence of the ``normalized'' buckling frequency, $\omega^2/\theta^4$ (red).
The dashed lines are power-law fits given by Eq.~(\ref{eq:S12}).
}
\label{fig:S5}
\end{figure}

For $\theta<\theta_c$, the buckling frequency can be extracted by small oscillations around equilibrium once the dynamics is started with $Q_0+\delta Q$ (with $\delta Q/ Q_0 \ll 1$).
For $\theta>\theta_c$, where the static buckling vanishes ($Q_0=0$),
the buckling frequency can still be extracted by injecting a small initial amplitude buckling distortion ($\Delta z_\mathrm{max} \ll d_0$) and then tracking the damped oscillations as the TBG evolves towards its unbuckled ground state .
Note that the frequency of the soft mode so injected, whose eigenvector has the wavelength of the  $(2 \times 1)$ buckling unit cell (i.e., 2 moir\'es), does not change as $N_\mathrm{moire}$ increases. As shown in Fig.~\ref{fig:S5}, the buckling frequency $\omega \to 0$ as $\theta \to \theta_c$.

The stiffness $K$ can be expressed as
\begin{equation}
    K= c \frac{\rho_\mathrm{2D}}{16 \pi^4} (\sqrt{3} a_\mathrm{Gr})^4  \frac{\omega^2}{\theta^4} \left(\frac{180}{\pi}\right)^4
    \label{eq:S11}
\end{equation}
where $c$ is a prefactor of $O(1)$. The last term is introduced so as to express the twist angle $\theta$ in degrees rather than radians. The twist dependence of $K$ clearly reflects that of $\omega^2/\theta^4$.
By fitting the simulation results (red circles in Fig.~\ref{fig:S5}b), we get:
\begin{equation}
    \frac{\omega^2}{\theta^4}=\left\{
    \begin{aligned}
    3.24\times 10^{21} (\theta_c - \theta)^{1.30}~\mathrm{s}^{-2} &, & (\theta<\theta_c)\\
    8.58\times 10^{20} (\theta - \theta_c)^{0.37}~\mathrm{s}^{-2} &, & (\theta>\theta_c)
    \end{aligned}
    \right.
    \label{eq:S12}
\end{equation}
With the twist angle dependence of $K$  thus obtained, the predictions of the zigzag model are:
(a) For $\theta \to \theta_c$, the buckling structure is infinitely soft ($K \to 0$), and it follows  that $D_\mathrm{eff} \to 0$. This agrees with the bending stiffness collapse discovered in bending simulations.
(b) On the low $\theta$ side $\theta<\theta_c$, with $D_\mathrm{f} \approx 100~\mathrm{eV} \gg K$, the effective bending stiffness $D_\mathrm{eff} \sim K$. According to Fig.~\ref{fig:S5}b, $K$ (and thus $D_\mathrm{eff}$) increases as the twist angle decreases from $\theta_c$, with an exponent $\gamma_{-}=1.30$ -- close to the bending simulation exponent $\epsilon_{-}=1.44$.
(c) On the large $\theta$ side $\theta>\theta_c$ with $D_\mathrm{f} \approx 2D_0 = 2.88~\mathrm{eV} \ll K$, the effective bending stiffness $D_\mathrm{eff} \sim 2D_0$. According to Fig.~\ref{fig:S8}b, $D_\mathrm{eff}$ decreases as the twist angle decreases approaching $\theta_c$, with an exponent $\epsilon_{+}= 0.2 $ -- not too far from the soft mode exponent $\gamma_{+}= 0.3(7) $.

\begin{figure}[ht!]
\centering
\includegraphics[width=0.8\linewidth]{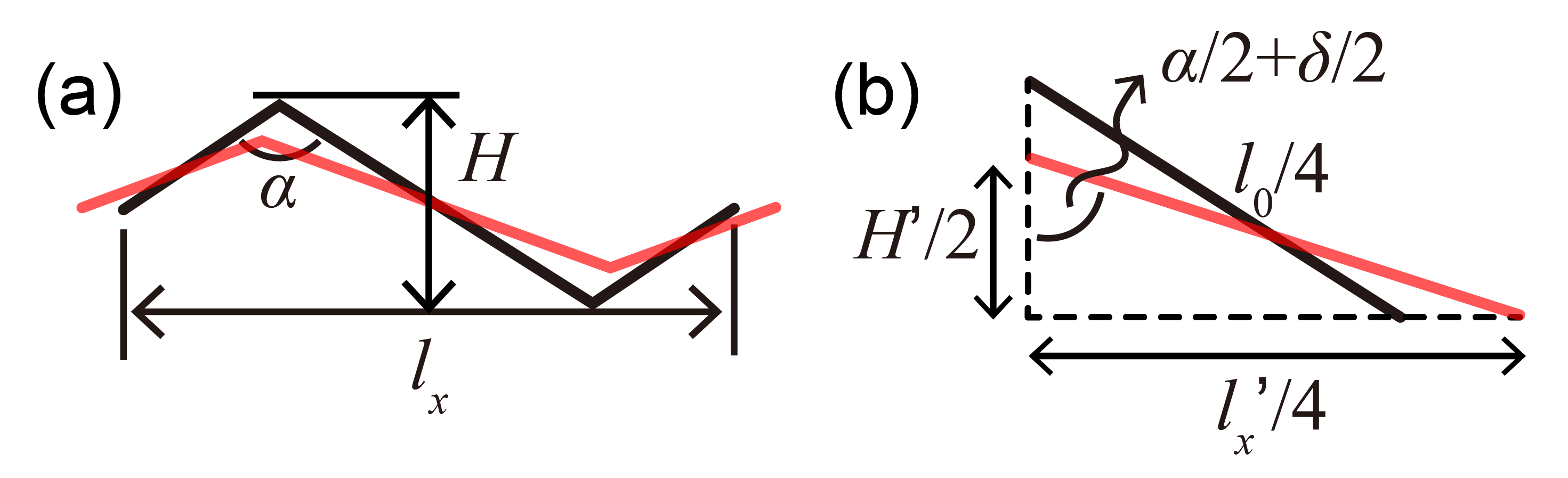}
\caption{(a) Model for elongation zigzag model, where black and red represent the equilibrium and elongated structure. (b) The quarter part of the model.}
\label{fig:S6}
\end{figure}

To fully address the magnitude of $K$ and compare theory and bending simulations, we need to specify the value of parameter $c$ in Eq.~\ref{eq:S11}. To achieve this, we performed additional ``elongation'' simulations.
The elongation simulations are constructed with quasi-static protocol -- stretching the size of the simulation box along $x$ direction by $\delta l_x = 0.002~\mathrm{\AA}$ ($\delta l_x \ll l_x$) in each step and performing the structural optimization.
The total elongation is $\Delta l_x = \delta l_x \times N_\mathrm{step}$, where $N_\mathrm{step}$ is the number of stretching steps.

The elongation simulation is sketched in Fig.~\ref{fig:S6}, where $l_0$ is the size of the $(2\times1)$ unbuckled structure (along $x$), $l_x$ and $H$
are the size and corrugation height of the buckled structure, $l_x'$ and $H'$ are the size and corrugation height during the elongation simulation, $\Delta l_x$ is the elongation, $\alpha$ is the equilibrium angle of each hinge and $\delta$ is the increase of the angle. These parameters are related by
\begin{equation}
\begin{aligned}
    H &=l_0/2 \cos(\alpha/2) \\
    l_x &=l_0 \sin(\alpha/2) \\
    H' &=l_0/2 \cos(\alpha/2+\delta/2) \\
    l_x' &=l_0 \sin(\alpha/2+\delta/2) \\
    \Delta l_x &= l_x'-l_x
    \label{eq:S13}
\end{aligned}
\end{equation}
where $\delta \ll \alpha$ and $\alpha+\delta \leq \pi$. From the elongation simulation, one can extract the tensile stiffness (of the buckled structure) $A$ by fitting the energy increase $\Delta E = \frac{1}{2} A \Delta l_x^2$.
Considering that the flat region can be regarded as a rigid body, this $\Delta E$ is mainly contributed by the increase of the hinge energy $2 \times \frac{1}{2} K \delta^2$, where the prefactor 2 represents two hinges in a $(2\times1)$ moir\'e unitcell.
Substituting the relationship
$\delta = \Delta l_x / H$ (which can be derived from Eqs.~\ref{eq:S13} above)
into $\Delta E$ and equating them, we finally have the hinge stiffness $K=A H^2/2$.
The value of $A$ and $K$ from elongation simulations is listed in Table S2. Comparing the value of $K$ in Table S2 with Eq.~\ref{eq:S11} and \ref{eq:S12}, we find that $c \approx 1.75$ gives the best match.

\begin{table}[ht]
\caption{\label{tab:table2} hinge stiffness of several buckled structures.}
\begin{ruledtabular}
\begin{tabular}{ccccc}
twist angle $\theta$ ($^\circ$) & 2.87 & 3.15 & 3.48 & 3.67\\
\hline
$A$ (N/m) & 39.12 & 41.36 & 47.66 & 46.13 \\
$H$ (\AA) & 2.958 & 2.272 & 1.323 & 0.6135 \\
$K$ (eV) & 10.69 & 6.67 & 2.61 & 0.543 \\
\end{tabular}
\end{ruledtabular}
\end{table}

\newpage
\section{Bending stiffness at finite temperature}
At finite temperature, flexural fluctuations immediately grow for our very large supercells, where the out-of-plane deformation of the buckling structure is heavily influenced by the thermal noise. Thus, it is difficult to use either the oscillation frequency method or the energy method to extract bending stiffness. We introduced an alternative method that is more effective at higher temperatures ($T\gtrsim T_c$), where all soft buckling phonon modes are largely populated.
Since we wish to extract the flexural phonon dispersion
(%and 
thus the bending stiffness) from MD simulations, we turn to the  power spectral density (PSD) method \cite{Thomas.prb.2010}.
The PSD method projects the velocity of each carbon atom to graphene phonon modes:
\begin{equation}
    P(\vec{k},\omega)=\frac{1}{4\pi \tau N} \sum_\alpha \sum_l m_l \Bigg|  \int_0^{\tau} \sum_{n=1}^N \Dot {u}_{l \alpha \vec{R}_n} (t) \exp \left( i \vec{k} \cdot \vec{R}_n -i\omega t \right ) \mathrm{d}t \Bigg|^2
    \label{eq:S15}
\end{equation}
where $\tau$ is the simulation time, $N$ is the total number of the primitive cell, ${u}_{l \alpha \vec{R}_n} (t)$ is the displacement of carbon atom in the $\alpha$ direction ($\alpha=x, y, z$) of atom $l$ (with mass $m_l$) inside primitive cell $\vec{R}_n$. Here $l$ and $n$ is the atom index inside one graphene primitive cell ($l=1, 2$) and the index of the primitive cell ($n=1,2,..., N$) respectively.\\

\begin{figure}[ht!]
\centering
\includegraphics[width=0.75\linewidth]{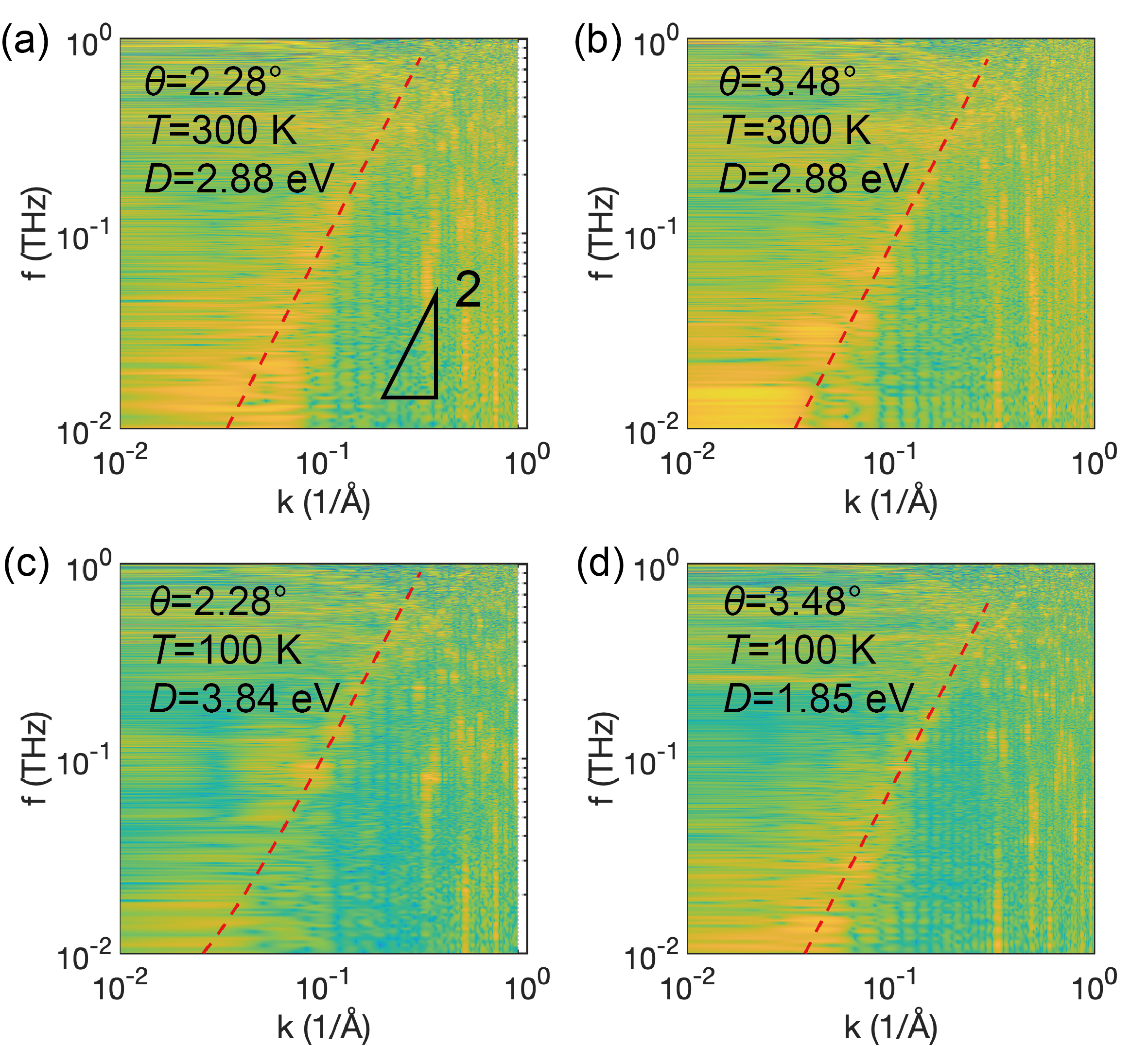}
\caption{Power spectral density of the upper layer of a twisted graphene bilayer at finite temperature. (a, b) Room temperature results for $\theta=2.28^\circ$ and $\theta=3.48^\circ$. (c, d) %Temperature 
$T=100$~K results for $\theta=2.28^\circ$ and $\theta=3.48^\circ$. Here the frequency is $f=\omega/2\pi$.
The spectrum for the lower layer is same as the upper layer.}
\label{fig:S8}
\end{figure}

In this way, the distribution of kinetic energy in each mode can be directly displayed, %that is, 
yielding
the dispersion relation of the system.
The dispersion relation of the flexural mode $\omega \propto k^2$, and the bending stiffness of the system can be obtained by fitting the prefactor
\begin{equation}
    D=\frac{\rho_\mathrm{2D} \omega^2}{k^4}
    \label{eq:S15}
\end{equation}
where $\rho_\mathrm{2D}$ is the mass of the bilayer per unit area. Several typical results of $\log_{10} [P(k,\omega)]$ (maxima used to identify the flexural mode) and the corresponding fit extracting the effective bending stiffness $D$ are shown in Fig.~\ref{fig:S8}.

For low temperature $T \ll T_c$,
the buckling order parameter $Q \approx Q_0$, thus we expect that the bending stiffness of the system is dominated by the ``hinge'' stiffness $K$, and the value of $D$ should be similar to $T=0$ cases.

\clearpage
\section{Tilt angle}

The buckled $(2 \times 1)$ bilayer normal locally deviates from $\vec{\hat{z}}$,  owing to the tilt of flat AB and BA regions.

\begin{figure}[ht!]
\centering
\includegraphics[width=0.9\linewidth]{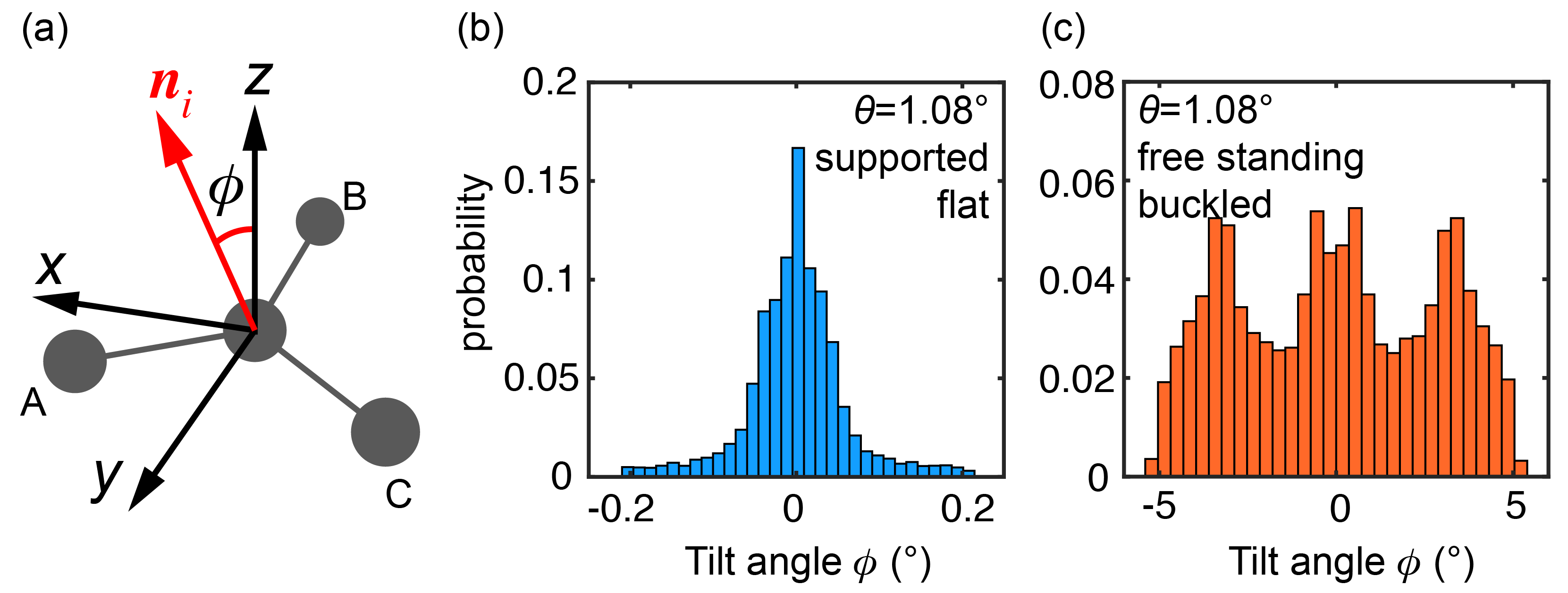}
\caption{(a) Schematic diagram of the local normal $\vec{n}_i$ and the tilt angle $\phi$. Probability distribution of $\phi$ for the magic angle (b) unbuckled and (c) buckled structures.}
\label{fig:S9}
\end{figure}

The tilt angle of $i$-th atom is defined as:
\begin{equation}
    \phi_i = \arcsin(\frac{\vec{n}_i \cdot \vec{e}_x}{|\vec{n}_i|})
    \label{eq:S16}
\end{equation}
where $\vec{n}_i$ is the local normal vector (Fig~\ref{fig:S9}a, with A, B and C the nearest neighbors of the central carbon atom).
Take $\theta_m=1.08^{\circ}$ TBG ($T=0$) as an example, the distribution of $\phi$ for supported-flat structure and freestanding-buckling structure are shown in Fig.~\ref{fig:S9}b and c respectively.
The maximum and the most probable tilt angles of the buckling structure ($\phi_\mathrm{max} \approx 5^{\circ}$ and $\phi_\mathrm{peak}\approx 3^{\circ}$) are more than an order of magnitude larger than the flat structure's ($\phi_\mathrm{max} \approx 0.2^{\circ}$). In addition, the distribution is also different, single peak for the flat structure, split peaks for buckled structure. Experiments detecting the tilt angle of the surface should easily capture the buckled structure when present.

\section{ $\sqrt{3} \times \sqrt{3}$ Buckling}

Besides the $2\times 1$ buckling described in main  text, energy minimization in larger simulation cells and zero planar stress found an alternative $\sqrt{3} \times \sqrt{3}$ buckled structure at only slightly higher total energy.

To convey a feeling for the relative difference, at $\theta$ = 3.15 degrees for example, the total energy per atom $E_\mathrm{tot}/ N_{at}$ is 
$E_i^{(\sqrt{3}\times \sqrt{3})}=-7.418088$~eV for $\sqrt{3} \times \sqrt{3}$ buckling, against $E_i^{(2\times 1)}=-7.418090$~eV for $2\times 1$ buckling and $E_i^\mathrm{flat}=-7.418076$~eV for no buckling.
These differences are not as imperceptible as they seem, once scaled with the large supercell size, and more importantly against the modest buckled-unbuckled entropy difference, resulting for example in an unbuckling temperature estimated to be as large as 65 K for the $(2\times 1)$ bilayer at this twist angle (Table S1).

It might of course happen that, owing to contingent factors such as imperfect structure caused by ill-fitting of $\sqrt{3} \times \sqrt{3}$ buckling inside the rectangular simulation box, force field inaccuracy, different entropies of the two phases at finite temperature, etc., the balance could be reversed, with $\sqrt{3} \times \sqrt{3}$
slightly more  stable than $2\times 1$. Even then, the former would remain an isolated possibility, whereas most uniaxial perturbations, intentional or accidental, would generally stabilize the latter.
For this additional reason we concentrated on the $2\times 1$ buckling.

\begin{figure}[ht!]
\centering
\includegraphics[width=0.88\linewidth]{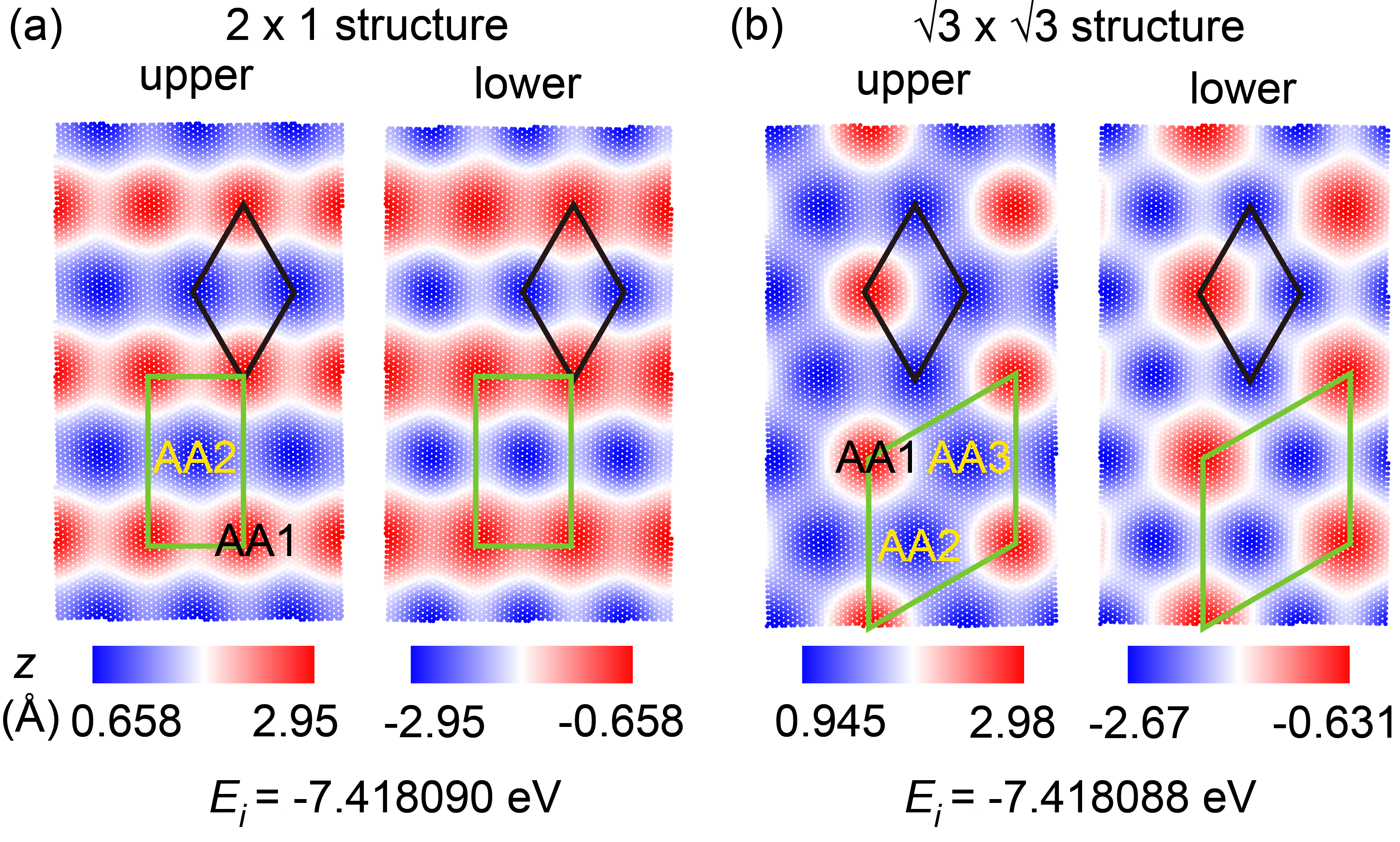}
\caption{
The $2\times 1$ and $\sqrt{3} \times \sqrt{3}$
buckling structures for $\theta=3.15^\circ$. The moir\'e primitive cell is highlighted by the black rhombus; the corresponding 2-moir\'e and 3-moir\'e cells and simulation boxes are shown with green rectangle/rhombus. The corrugation height and the
potential energy $E_i$ per atom are listed at the bottom.}
\label{fig:Sfolded}
\end{figure}

The detailed geometry of the two buckled structures at the same twist angle is shown in Fig.~\ref{fig:Sfolded}.
In the $\sqrt{3} \times \sqrt{3}$ distortion, where $C_{3z}$ symmetry is fully restored, the up-down symmetry is broken, with a triangular AA1 superlattice strongly raised out of plane, and a complementary honeycomb superlattice comprising two weakly lowered AA2 and AA3 per cell (Fig.~\ref{fig:Sfolded}b). Further below we will show in Fig.~\ref{3fold}, for completeness, the 12 electronic bands of the $\sqrt{3}\times \sqrt{3}$ buckled structure, of course different from those of $2\times 1$.

\clearpage
\section{Electronic structure}

\subsection{Hamiltonian}

The tight binding Hamiltonian is
$H = \sum_{i,j} -t_{ij} |i \rangle \langle j|  + {\rm H.c.}$,
where $t_{ij}$ is the transfer integral between sites $i$ and $j$. Let $\boldsymbol{R}_j - \boldsymbol{R}_i = (l,m,n) d$ be the distance vector between two orbital, and $\boldsymbol{\hat{n}}_i = (n_i^l,n_i^m, n_i^n)$ be the direction of orbital $\boldsymbol{p}_i$, then using the Slater-Koster\cite{slater1954LCAO} formula
\begin{equation}
\begin{split}
    t_{ij}= \sum_{k}^{l,m,n} n_{i}^{k}  \Big[ n_{j}^{k} (k^2 V_{pp\sigma} +(1 - k^2 )V_{pp\pi} ) + 
    \sum_{g \neq k}^{l,m,n} n_{j}^{g} gk(  V_{pp\sigma} -V_{pp\pi}) \Big] 
    \label{eq:tijTrue}
\end{split}
\end{equation}
where the out-of-plane $(\sigma)$ and in-plane $(\pi)$ hoppings are
\begin{equation}
    V_{pp\sigma} = V_{pp\sigma}^0 \; \exp{\left(-\frac{d-d_0}{r_0}\right)}, \;\;\; V_{pp\pi} = V_{pp\pi}^0 \; \exp{\left(-\frac{d-a_0}{r_0}\right)}
\end{equation}
where $V_{pp\pi}^0=-2.7$~eV and $V_{pp\sigma}^0=0.48$~eV are are chosen to reproduce ab-initio curves in AA and AB stacked bilayer graphene similar to previous works \cite{Mattia-PRX2019}, $d_0=3.344~\mathrm{\AA}$ is the minimum distance between the two layers in the AB stacked region, $a_0 = a_\mathrm{Gr} / \sqrt{3} \approx 1.42~\mathrm{\AA}$ is the carbon-carbon bond length in the relaxed structure, and $r_0$ is the decay rate $\approx 0.184 a_\mathrm{Gr}$.

In the particular case of unbuckled, flat bilayer $\boldsymbol{\hat{n}}_i \approx \boldsymbol{\hat{n}}_j \approx (0,0, 1)$ therefore Eq. \eqref{eq:tijTrue} reduces to $t_{ij}=  {n}^2 V_{pp\sigma}  + (1-n^2)  V_{pp\pi} $ 

\begin{figure*}[h]
\centering
\includegraphics[width=1\linewidth]{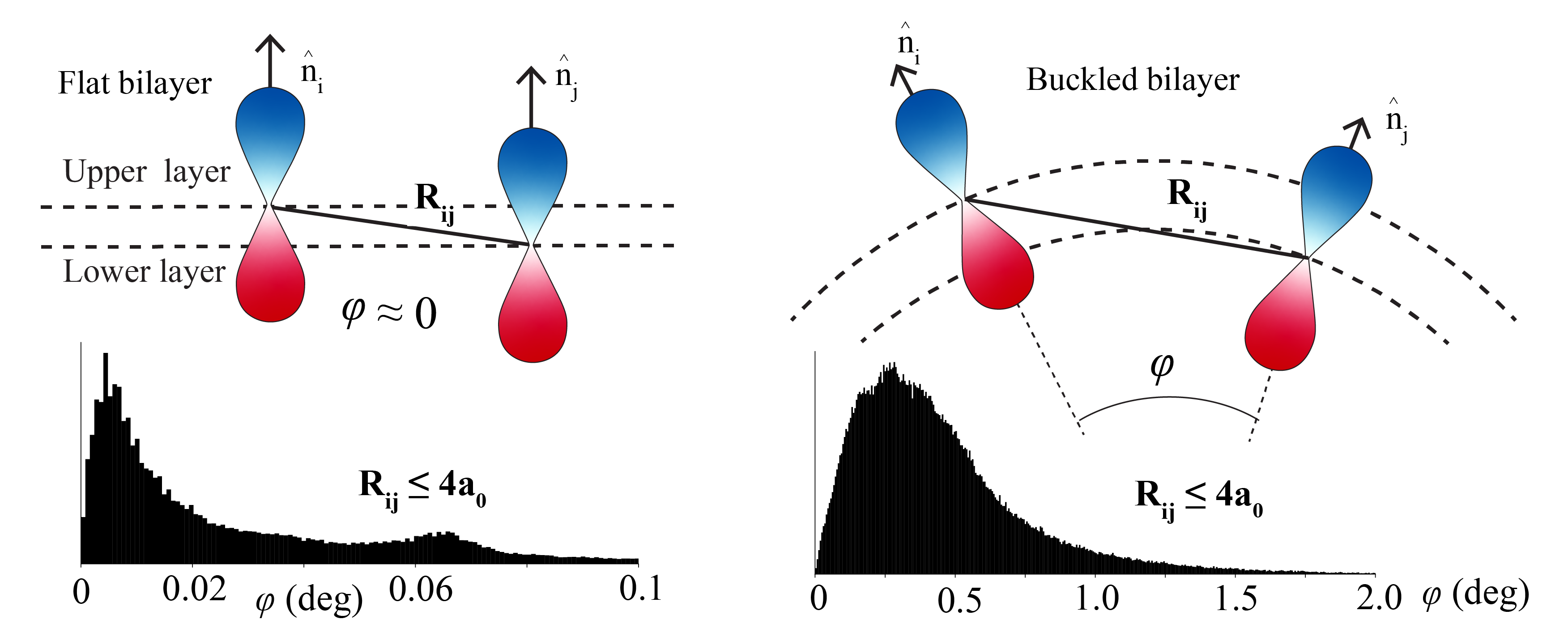}
\caption{Unlike the flat bilayer (left panel), where local normals deviate very little from vertical, the deviation is  much larger in the buckled state (right panel).
The distribution of the included angle $\varphi$ between two local normals $\hat{n}_i$ and $\hat{n}_j$ (within the cutoff range $R_{ij} \leq 4 a_0$) is shown in the lower panel.
The maximum $\varphi$ for the buckled structure is more than one order of magnitude larger than the unbuckled structure -- similar to what we discovered for the tilt angle $\phi$.
}
\end{figure*}

\clearpage
\subsection{Space group}

\begin{figure}[t]
\centering
\includegraphics[width=0.75\linewidth]{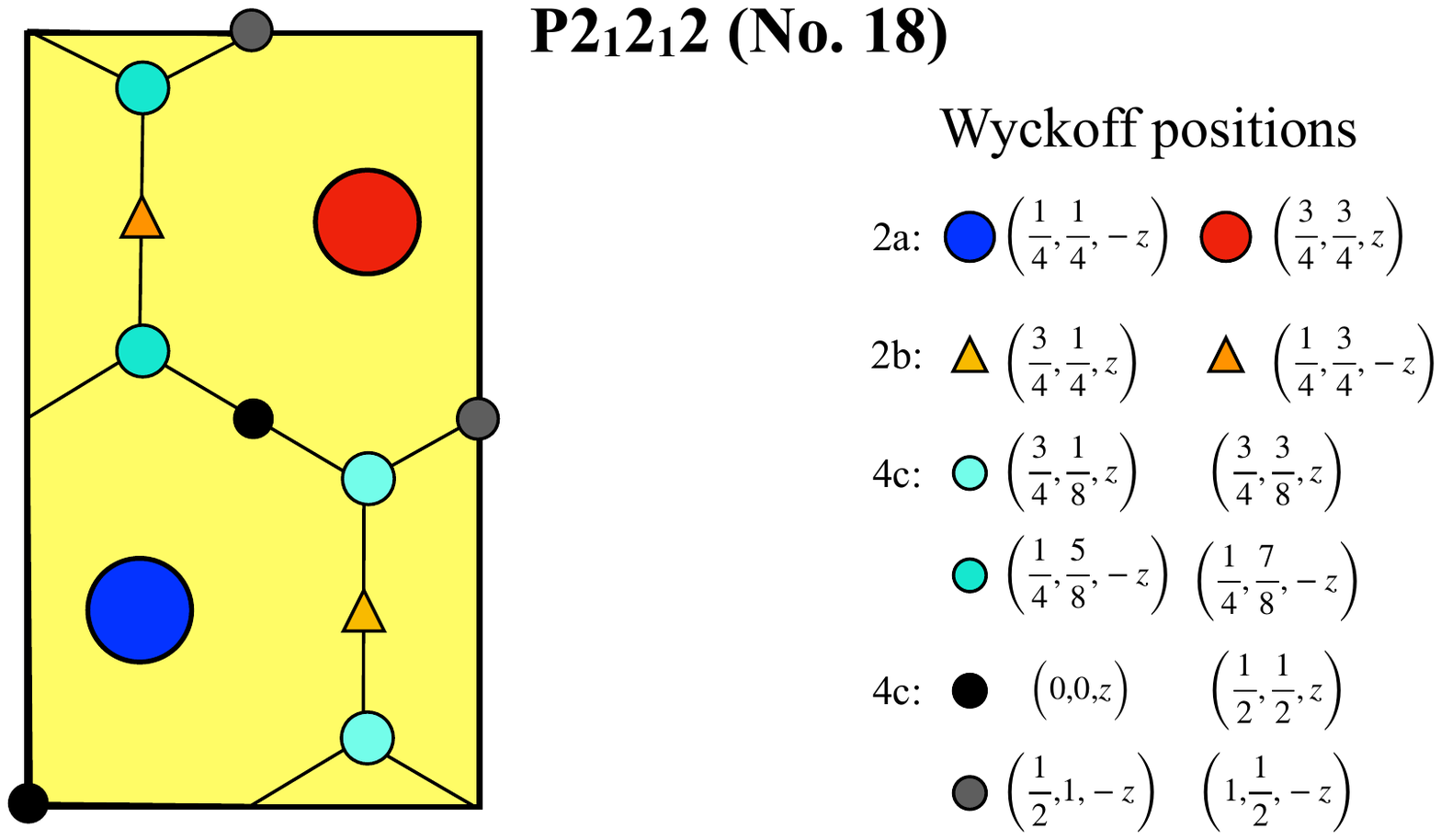}
\caption{Unit cell and Wyckoff positions in the space group $P2_12_12$. Note that the 2a positions corresponds to the AA stacked regions, whereas the 4c positions in cyan to the Bernal stacked ones.}
\label{unit-cell}
\end{figure}

The unit cell of the $(2 \times 1)$ buckled TBG together with the Wyckoff positions in the $P2_12_12$ space group (no. 18) is shown in Fig.~\ref{unit-cell}. We note that the 2a Wyckoff positions (red/blue circles) correspond to the AA stacked regions and the 4c positions (cyan circles) to the Bernal stacked ones. The 3f positions in the P622 space group (no. 177) of the unbuckled TBG split into the 2b and 4c (yellow triangles and black circles, respectively) in the space group of the buckled TBG. The non-symmorphic symmetry connects the two 2a positions, the two 2b ones, as well as the two sets with opposite $z$ coordinates within the 4c Wyckoff positions (light cyan $\to$ dark cyan, as well as dark gray $\to$ black).

\subsection{Symmetries}

\begin{table}[h]
\bealn
&
\begin{array}{|c|c|c|c|c|}\hline
~ & E & C_{2z} & C_{2y} & C_{2x}\\ \hline
\Gamma_1 & 1 & 1 & 1 & 1 \\ \hline
\Gamma_2 & 1 & 1 & -1 & -1 \\ \hline
\Gamma_3 & 1 & -1 & -1 & 1 \\ \hline
\Gamma_4 & 1 & -1 & 1 & -1 \\ \hline
\end{array}
&
&
\begin{array}{|c|c|c|c|c|}\hline
~ & E & C_{2z} & C_{2y} & C_{2x}\\ \hline
W_1 & 1 & 1 & i & i \\ \hline
W_2 & 1 & 1 & -i & -i \\ \hline
W_3 & 1 & -1 & -i & i \\ \hline
W_4 & 1 & -1 & i & -i \\ \hline
\end{array}&
&
\begin{array}{|c|c|c|c|c|}\hline
~ & E & C_{2z} & C_{2y} & C_{2x}\\ \hline
X_1(2) & 1 & \sigma_3 & \sigma_1 & -i\sigma_2 \\ \hline
\hline
Y_1(2) & 1 & \sigma_3 & -i\sigma_2 & \sigma_1\\ \hline
\end{array}
\eal
\caption{Character table and irreducible representations at the high-symmetry points in space group $P2_12_12$.}
\label{Tab-character-table}
\end{table}

\begin{figure}[h!]
\centering
\includegraphics[width=0.9\linewidth]{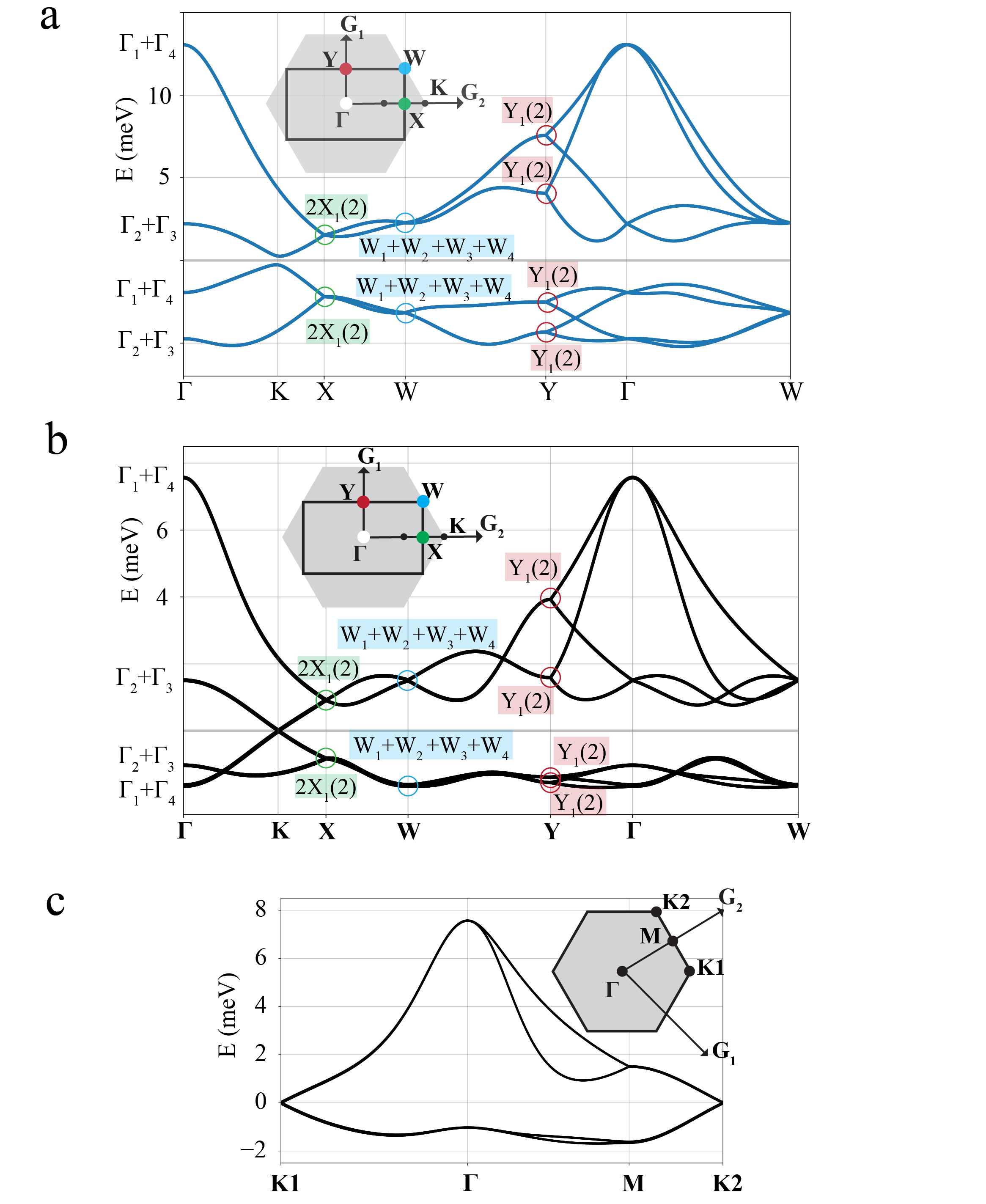}
\caption{
Flat bands for the bilayer graphene with twist angle $\theta =1.08^{\circ}$. Top panel (a): flat bands of the $(2 \times 1)$ buckled bilayer; middle panel (b): flat bands of the unbuckled bilayer folded into the $(2 \times 1)$ reduced Brillouin Zone. The additional crossings of the unbuckled bands arise because the actual space group is P622 (no. 177). Bottom panel (c): flat bands of the unbuckled bilayer in their natural (1 $\times$ 1)  Brillouin Zone. }
\label{flat-bands}
\end{figure}

% \begin{table}[h]
% \bealn
% &
% \begin{array}{|c|c|c|c|}\hline
% ~ & C_{2z} & C_{2y} & C_{2x}\\ \hline
% \Gamma_1 & 1 & 1 & 1 \\ \hline
% \Gamma_2 & 1 & -1 & -1 \\ \hline
% \Gamma_3 & -1 & -1 & 1 \\ \hline
% \Gamma_4 & -1 & 1 & -1 \\ \hline
% \end{array}
% &
% &
% \begin{array}{|c|c|c|c|}\hline
% ~ & C_{2z} & C_{2y} & C_{2x}\\ \hline
% W_1 & 1 & i & i \\ \hline
% W_2 & 1 & -i & -i \\ \hline
% W_3 & -1 & -i & i \\ \hline
% W_4 & -1 & i & -i \\ \hline
% \end{array}&
% &
% \begin{array}{|c|c|c|c|}\hline
% ~ & C_{2z} & C_{2y} & C_{2x}\\ \hline
% X_1(2) & \sigma_3 & \sigma_1 & -i\sigma_2 \\ \hline
% \hline
% Y_1(2) & \sigma_3 & -i\sigma_2 & \sigma_1\\ \hline
% \end{array}
% \eal
% \caption{Character table and irreducible representations at the high-symmetry points in space group $P2_12_12$.}
% \label{Tab-character-table}
% \end{table}
In Table~\ref{Tab-character-table} we report the character table of the irreducible representations (irreps) at the high-symmetry points in space group $P2_12_12$, where $\sigma_a$, $a=1,2,3$, are Pauli matrices that act in the space of two-dimensional irreps. The eigenstates of the Hamiltonian that we calculate do transform as one of the irreps at the corresponding high-symmetry points, see Fig.~\ref{flat-bands}(a), supporting  our identification of the space group.

In Fig.~\ref{flat-bands}(b) we also show the flat bands of the \textit{unbuckled} bilayer folded into the reduced BZ of the $2 \times 1$ buckled one, labelling the Bloch waves at the high symmetry point with the irreps of the $P2_12_12$ space group, (despite the correct space group in this case being $P622$, no. 177). We remark that, proceeding as outlined in the main text, in this unbuckled case too we can identify the flat bands corresponding to a given valley, magnetic space group $P2_12'_12'$, no. 18.19. Like in the buckled case, these single-valley bands can be generated by the atomic limit of Wannier orbitals centered at the Wyckoff positions $4c$ of space group $P2_12'_12'$. Yet, this was not possible in the original unfolded band structure~\cite{Bernevig-PRL2019, Mattia-PRX2019}: and that was a clear symptom of non-trivial topological character of the unbuckled bilayer -- {\it a property that should be independent of the arbitrary choice of unit cell and folding procedure}. For that reason, despite our inability to prove it, we conjecture that the bands of the buckled twisted bilayer graphene are likely to be topological as well.\\

\clearpage

\begin{figure}[h!]
\centering
\includegraphics[width=0.8\linewidth]{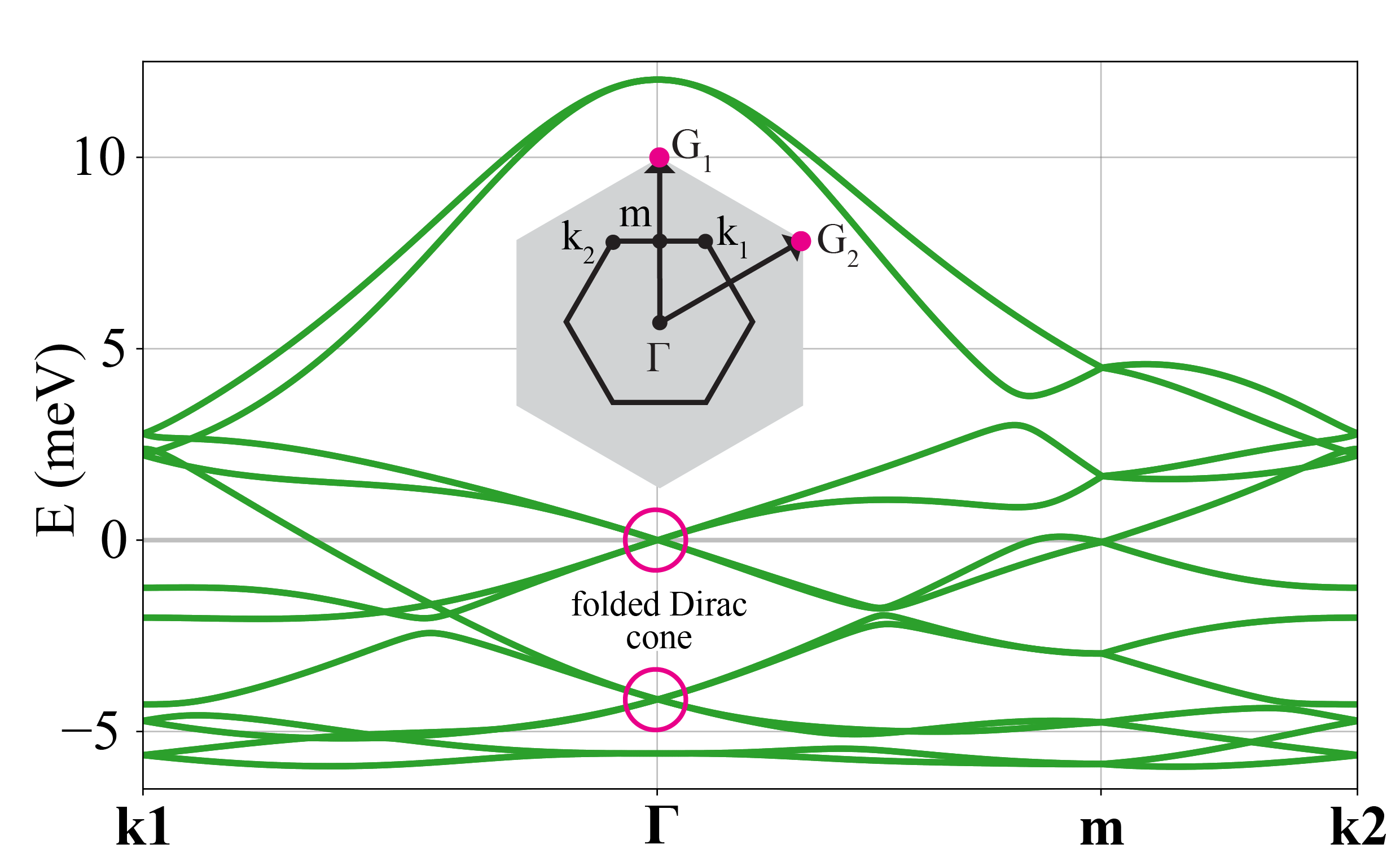}
\caption{Flat bands for the $\sqrt{3} \times \sqrt{3}$
buckled bilayer with twist angle $\theta =1.08^{\circ}$. Note the double folding of the Dirac cone at $\Gamma$. Beware of the 90 degrees rotated convention compared with that of $(2 \times 1)$ bands of other figures.}
\label{3fold}
\end{figure}

For completeness, the flat bands of the $\sqrt{3}\times \sqrt{3}$ buckled structure are shown in Fig.~\ref{3fold}.

\subsection{\texttt{LTB+Symm} code}

For this work, we developed the \texttt{LTB+Symm} code (Large-scale Tight Binding and Symmetry validation). \texttt{LTB+Symm} is a useful tool to perform tight binding calculations with non-planar geometries in large-sized unit cells, thanks to its MPI implementation. \texttt{LTB+Symm} is an object-oriented, open-source Python3 software; 
 its streamlined, geometry-agnostic implementation allows to easily implement \textit{ad-hoc} solutions to capture system-specific features. A novel aspect  of \texttt{LTB+Symm} is its ability to validate and label irreps at high-symmetry points of the Brillouin zone (once the right space group is defined).
\texttt{LTB+Symm} is publicly accessible on GitHub:  https://github.com/khsrali/LTB-Symm.

\newpage
% \section*{References}
\bibliographystyle{unsrt}
\bibliography{ref}